\def\lae{\mathrel{<\kern-1.0em\lower0.9ex\hbox{$\sim$}}}

\documentclass[12pt,preprint]{aastex}

\newcommand{\gae}{\mathrel{>\kern-1.0em\lower0.9ex\hbox{$\sim$}}}

\lefthead{Boroson et al.}
\righthead{Sco X-1 Observed with {\it FUSE}}

\begin{document}

\title{FUSE Observations of a Full Orbit of Scorpius X-1}

\author{Bram Boroson}
\affil{Clayton State University
2000 Clayton State Blvd
Morrow, GA 30260
bramboroson@clayton.edu}

\and

\author{Saeqa Dil Vrtilek and John Raymond}
\affil{Smithsonian Astrophysical Observatory
Mail Stop 83
Cambridge, MA 02138
svrtilek@cfa.harvard.edu, jraymond@cfa.harvard.edu}


\begin{abstract}

We obtained UV spectra of X-ray binary Scorpius X-1 in the 900--1200\AA\ 
range with the Far Ultraviolet Spectroscopic Explorer over the 
full 0.79~day binary orbit. The strongest emission lines are the doublet 
of O\,VI at 1032,1038\AA\ and the C\,III complex at 1175\AA. The spectrum 
is affected by a multitude of narrow interstellar absorption lines, both 
atomic and molecular. Examination of line variability and Doppler 
tomograms suggests emission from both the neighborhood of the donor star 
and the accretion disk. Models of turbulence and Doppler broadened 
Keplerian disk lines Doppler shifted with the orbit of the neutron star 
added to narrow Gaussian emission lines with 
undetermined Doppler shift fit the data 
with consistent values of disk radius, inclination, 
and radial line brightness profile. The Doppler shift of the narrow 
component with the orbit suggests an association with the donor star. We 
test our line models with previously analyzed near UV 
spectra obtained with the Hubble Space Telescope Goddard High Resolution 
Spectrograph and archival spectra obtained with the HST Cosmic 
Origins 
Spectrograph.

\end{abstract}


\section{Introduction}

Sco~X-1 is prototype for the ``Z sources" that trace out Z-shaped tracks 
in X-ray color-color diagrams. These neutron stars accreting from low mass 
companions typically do not show X-ray pulsations, but some may display 
instead bursts due to runaway nuclear burning of accreted gas, confirming 
the identification of the compact objects as neutron stars with the 
measurement of radii from the burst blackbody spectra.

Searches for X-ray periodicities have instead revealed multiple broad 
power spectral peaks from $\sim5-1000$~Hz, the Quasi-Periodic 
Oscillations, as well as a complex and variable noise continuum. The 
neutron stars in the Z~sources and similar atoll sources may have lower 
magnetic fields than the neutron stars in pulsars, leading the inner 
accretion disk to penetrate closer to the surface of the neutron star. The 
kHz QPO phenomenon probably results from some interaction of a rapidly 
spinning neutron star and the rapid rotation of the inner accretion disk. 
Models of the QPOs promise to measure the mass and radius of the neutron 
star (and thus constrain the equation of state of the degenerate interior) 
but different QPO models lead to different results.

Studies of spectral variability could help to confirm or refute some of the QPO models. For example, the
mass of the neutron star could be bounded by orbital Doppler shifts from the disk and donor star, limits
on the inclination, or through Keplerian velocities in the accretion disk.
Sco~X-1 is close (2.8$\pm0.3$~kpc from parallax measurements with the 
VLBA, Bradshaw, Fomalont
\&\ Geldzahler 1999) and as a Z~source is expected to emit near the Eddington limit for a
neutron star ($\sim10^{38}$~erg~s$^{-1}$). Yet
the accretion kinematics and system parameters are not as well known as for 
Cataclysmic Variables (CVs), more common (and thus closer) accreting white 
dwarf 
systems,
or for X-ray binaries with pulsars or with a high enough inclination for eclipses and X-ray dips.
We still remain ignorant of the shape of the accretion disk,
whether it is warped and precesses, and the role, if any, of winds from the disk and companion star that
might be caused by the X-ray illumination.

The optical counterpart was found by Sandage et al. (1966), and the orbital period of 0.787 days is detected
in optical variability (Gottlieb, Wright, \&\ Liller 1975). The optical spectrum includes variable lines
of \ion{He}{2}$\lambda4686$ (Cowley \&\ Crampton 1975) and a complex of lines from \ion{N}{3} and
\ion{C}{3} near 4640\AA\ which result from the Bowen fluorescence process 
in which He\,II Lyman~$\alpha$
pumps EUV O\,III lines which in turn excite N\,III (Bowen, 1934, 1935;
Willis et 
al. 1980; Schachter, Filippenko,
\&\ Kahn 1989).

The UV waveband is particularly well suited for the study of the accretion 
disks in X-ray binaries, as the multiple resonance lines, with greater 
flux than the optical lines, can be used to diagnose the gas density, 
temperature, ionization stage, and optical depth. Vrtilek et al. (1991) 
used simultaneous {\it GINGA} and IUE observations to relate the X-ray 
state to the UV line and continuum strength. Kallman, Boroson, \&\ Vrtilek 
(1998) used the Goddard High Resolution Spectrograph on the Hubble Space 
Telescope to study the UV lines (with resolution up to $R\sim 25,000$) and 
continuum and perform cross-correlations with X-rays observed 
simultaneously with the Rossi X-ray Timing Explorer.

Sco X-1 has not been previously observed in the far UV (900-1200\AA). This 
wavelength range is of particular interest because of the \ion{O}{6} 
doublet at 1031.9 and 1037.6\AA. Along with the \ion{N}{5} and \ion{C}{4} 
doublets in the near UV that can be seen with HST (Kallman, Boroson, \&\ 
Vrtilek 1998), these are the strongest line signatures from X-ray 
ionization of the accretion disk. The \ion{O}{6} ion has a slightly higher 
ionization potential than the near UV ions, and a wider doublet separation 
which rules out overlap between the line components for most velocities 
expected from the accretion disk. Other important lines in this wavelength 
region include a blend of \ion{C}{3} at 1175\AA\ and an \ion{N}{3} line at 
991\AA\ that is involved in the Bowen fluorescence process that causes the 
4640\AA\ lines.

Reconstruction techniques such as Doppler tomography, which has been 
applied to CVs, and echo mapping, which has been applied to both CVs and 
AGN, have also been applied to Sco~X-1
(Steeghs \&\ Casares 2002, Mu\~noz-Daries et al. 2007).
These studies resulted in the first spectroscopic measurement of the 
motion of the donor star and a limit on the mass ratio (Steeghs \&\ 
Casares 2002).

In order to investigate for the first time the far UV spectral lines, and to extend the mapping techniques to 
infer the system and accretion kinematics, we used the Far Ultraviolet Spectroscopic Explorer (FUSE) to obtain 
the spectra at times evenly spaced throughout the binary orbit. This time 
coverage allows us to construct Doppler 
tomograms and model orbital variations.

We have presented an analysis of the far UV spectrum of the X-ray binary 
Hercules~X-1 as recorded over a binary orbit (Boroson et al. 2007). The 
methods we applied in that case are generally 
applicable to Sco X-1 as well.  Since Sco X-1 does not eclipse we cannot apply eclipse-mapping 
techniques but we can apply tomography. 
Unfortunately, Sco~X-1 is also 
behind a much greater column density of interstellar gas, which we have to model in order to 
reach conclusions about the variable emission from the system itself.

This paper is organized as follows. In Section~2, we describe the observations in detail. In Section~3, we 
analyze the extensive signatures of interstellar absorption on the spectrum, including absorption lines from 
both atoms and molecular hydrogen. Accounting for the interstellar absorption is neccessary for accurate 
modeling of the intrinsic source spectrum. In Section~4, we describe the Doppler tomography of the lines, and in 
Section~5 we describe more detailed models of the emission line variability. In Section~6, we model the 
continuum.

\section{FUSE Observations}

{\it FUSE} was a NASA {\it Origins} mission operated by The Johns Hopkins
University. The spectral coverage is between 900 and 1190\AA, with 
a resolution of $\approx0.05$\AA. Four aligned telescopes feed two identical 
far-UV
spectrographs with resolution R$=20000$.
The FUSE mission is described in more detail in Moos et
al. (2000) and its on-orbit performance is described in Sahnow et  al.
(2000). 

Our {\it FUSE} observations began on August 19, 2001 at 5:59~UT. Table~1 
shows the log of exposures, each of which is integrated over each {\it 
FUSE} orbit of the Earth, with gaps when Sco~X-1 goes below the horizon. 
We use the orbital ephemeris of Steeghs \&\ Casares (2002) to determine 
the orbital phases (in the Sco~X-1 system) of our observation, which we 
also present in Table~1. To convert to orbital phases according to the 
updated ephemeris of Galloway et al. (2014), subtract 0.011 from the 
phases displayed. 
We supplement the FUSE observations with 
observations using the Cosmic Origins Spectrograph (COS) on the Hubble 
Space Telescope in 2010. The log of these observations is shown in 
Table~2.

We used the CalFUSE pipeline software version 3.2.1 to extract and
calibrate the data from all four {\it FUSE} telescopes.  Below 1100\AA, 
where emission features are sharp, we added an
offset to each wavelength scale, in intervals of 0.025\AA, so that the
absorption and emission features from each detector best agreed. 
We tested our wavelength calibration against the interstellar 
Si\,II$\lambda1020.6989$ absorption line, which we found to have a mean
heliocentric velocity of $-30$~km~s$^{-1}$. The standard deviation of 
the centroid of this line from orbit to orbit was $\approx0.03$\AA\  or 
$<10$~km~s$^{-1}$.

The S/N of the data was $\approx5$ per 0.1\AA\ pixel in the continuum in the
region $<1000$\AA\ and $\approx10$ near 1100\AA. The S/N within the O\,VI
line had greater variation with orbital phase, as the doublet changed both
in shape and in strength. At $\phi=0.75$, the peak S/N within the doublet
was $\approx15$ per 0.1\AA\ pixel, while at $\phi=0.5$, the peak S/N was
$\approx25$ per pixel. 

In Figure~1, we show the average observed FUSE spectrum and
compare it with the continuum model we developed based on near-UV 
observations (Vrtilek et al. 1991). We
indicate the same prominent stellar emission lines in the far UV as Robert
et al. (2003): S\,IV at 1062.7\AA,1073.0\AA, and 1073.5\AA,
C\,III multiplet at 1175.6\AA, P\,V at 1128.0\AA, 
Si\,IV
at 1128.3 \AA, and the O\,VI doublet at 1031.9 and 1037.6\AA. 

We mark geocoronal (airglow) lines produced in the Earth's atmosphere as 
given by Feldman et al. 
(2001): H\,I\,Lyman-$\gamma$ at 973\AA, O\,I at 989\AA, and 
H\,I\,Lyman-$\beta$+O\,I
from 1026-1027\AA.

\section{COS observations}

Spectra of Scorpius X-1 in the 1300--1450\AA\ range were taken in October 
22, 2010 using the Cosmic Origins Spectrograph (COS, Osterman et al. 2011) 
aboard the Hubble Space Telescope. 
We present a 
time-averaged spectrum, binned by 9 pixels, in Figure~\ref{fig:cos}. While 
the UV lines are known to vary with the X-ray state of the neutron star 
(Vrtilek et al. 1991), spectra with this high resolution and sensitivity 
may help constrain the models we develop to understand the FUSE spectra.

\section{Interstellar absorption features\label{sec:ism}}

We fit the H$_2$ lines to a broad range of the spectrum and only fit 
individual atomic absorption lines in the vicinity of the O\,VI doublet. 
Precise modeling of the interstellar absorption lines is necessary for 
interpretation of the emission features intrinsic to Sco~X-1.

To fit the molecular lines, we use line templates from McCandliss (2003), which are based
on Abgrall et al. (1993a, 1993b). We include all rotational states from 
$j=0$ to $j=6$, and allow two independent absorption components. Each 
component has its own fixed velocity width parameter and centroid offset, with 
column densities free from $j=0$ to $j=6$. The continuum is assumed to be 
linear, and only the wavelength ranges from 1000\AA to 1020\AA, 
1041\AA to 1048\AA, 1050\AA to 1071\AA, and 1080\AA\ to 1120\AA are used.
These wavelength ranges were chosen to exclude prominent emission lines,
which might vary and bias the measurement of interstellar absorption features.

The best fit total column density ($j=0$ to $j=6$) is
6.7$\times10^{19}$~cm$^{-2}$ and 1.0$\times10^{19}$~cm$^{-2}$ in the two 
components. Using $j=0$ and $j=1$ levels with statistical weights 
$g_1/g_0=9$ and the energy difference $\Delta E=170.5$~K, we find
the first component is at $\approx90$~K, consistent with H$_2$ towards
disk stars, whereas the second, weaker component is consistent with
the high temperature limit (Gillmon et al. 2006).

The saturated interstellar H Lyman$\beta$ line at 1026.7\AA\ provides
a measure of interstellar absorption, and its placement near the
O\,VI doublet requires that the O\,VI emission lines and Ly$\beta$ line be 
modeled 
together. We describe our model of the emission lines (from both 
accretion disk and donor star) in \S\ref{sec:lines}.

The Ly$\beta$ line was fit to a Voigt profile and was found to have a 
column 
density N$_{\rm H}=2.2\times10^{21}$~cm$^{-2}$. Kallman, Boroson, \&\ 
Vrtilek (1998) found N$_{\rm H}=8.9\times10^{20}$~cm$^{-2}$ from the 
Ly$\alpha$ line at 1216\AA, observed with the GHRS. Kallman, Raymond, \&\ 
Vrtilek (1991) found N$_{\rm H}=6\pm2\times10^{21}$~cm$^{-2}$ from fits to 
the Ly$\alpha$ absorption observed with {\it IUE}. 
Scaling from the 
H absorption to the expected reddening using the relation of 
Diplas \&\ Savage (1994), N$_{\rm 
H}/$E$(B-V)=4.93\times10^{21}$~cm$^{-2}$~mag$^{-1}$, we find 
E$(B-V)=0.45$. However, we note that there are deviations from the 
relation between E$(B-V)$ and N$_{\rm H}$.

\section{Doppler tomography}
 
The Doppler tomography method, developed by Marsh and Horne (Marsh
\&\ Horne 1988, Marsh 2005) takes as input an emission line which is 
broadened by line of sight motion. The line must be observed over a good 
sampling of the binary orbit.

We determine the tomograms using
Fourier-Filtered Back Projection. This involves applying the Fourier
transform to each spectrum, multiplying by a ramp filter and a Wiener
filter based on the noise level, and then taking the inverse Fourier
transform. The resulting spectra were then back-projected, or integrated,
to provide a tomogram. We did not apply  Maximum
Entropy regularization. 

As a test that our filtering technique does not introduce artifacts, we 
have obtained (Steeghs, D., personal communication) the optical spectra of 
Sco X-1 presented in Steeghs \&\ Casares (2002). Our tomograms of the 
optical lines (presented in Figures~\ref{opticaltomo}) give similar 
results to the Maximum Entropy Method used by Steeghs \&\ Casares for the 
lines that we both analyzed. We have assumed a systemic velocity of 
$-113.8$~km~s$^{-1}$ for the Sco~X-1 system (Steeghs \&\ Casares 2002).

The Doppler tomograms are limited by our phase coverage (15 time-separated 
spectra) and by the reconstruction of the underlying spectrum by 
correction for interstellar absorption lines (\S\ref{sec:ism}), but the 
exercise is valuable in rendering the system independent of particular 
models.

In Figure~\ref{fig:tomo} we show a Doppler tomogram of the blue component 
of 
the O\,VI doublet, which is less affected by interstellar absorption. We 
have interpolated between orbital phases for the integration, which we 
perform between $\phi=0.02$ and $\phi=0.98$. A compact bright spot on the 
tomogram overlaps with emission expected from the Roche lobe of the donor 
star. There is also a suggestion of an incomplete ring surrounding the 
position of the neutron star in the tomogram.

The continuum model developed prior to the observations was scaled to the 
continuum at each FUSE orbit and was subtracted from each spectrum. Where 
strong interstellar absorption lines of molecular hydrogen were present, 
we have adjusted for their presence. We fit the H$_2$ absorbers to the 
entire FUSE spectrum (section~\ref{sec:ism}), and divide the spectrum by 
this model. In the vicinity of the O\,VI doublet, we correct also for 
atomic  Lyman~$\beta$ absorption, using a simultaneous model for the 
emission and absorption lines (see section~\ref{sec:lines}). Thus the 
tomograms near O\,VI corrected for interstellar lines are not entirely 
model-independent. We interpolate over any resulting data points that have 
$<2\sigma$ significance to avoid magnifying the uncertain flux near the 
trough of saturated absorption lines. There may be some remaining 
unidentified and uncorrected interstellar features that overlap with 
O\,VI. There may be absorption from interstellar 
O\,VI itself, perhaps enhanced by photoionization of the surrounding ISM
by Sco~X-1 (McCray, Wright, \&\ Hatchett, 1977), but we could not identify
such lines unambiguously.

\section{Detailed Line Models\label{sec:lines}}

In addition to Doppler tomography, which makes no assumptions about the 
location of the emission, we also fit to the emission lines detailed 
empirical models that make use of what is known from other studies.

As a preliminary, we fit the lines with a single broad emission component 
(with a Gaussian or Lorentzian shape), subject to interstellar absorption, 
at each orbital phase. The centroid of the lines matched closely the 
expected orbital velocity of the disk (Steeghs \&\ Casares 2002). The 
tomograms suggest that emission is also present from a more compact 
region. However, models with line width, centroid, and normalization of 
both narrow and broad components left as free parameters were not 
constrained by the data.

As further evidence that the lines contain a component emitted in the disk 
we show in Figure~\ref{fig:lineav} that the average O\,VI line near 
$\phi=0.25,0.50,0.75$ displays a trend of Doppler shift as expected from 
the neutron star. Looking at the differences between the line spectra at 
$\phi=0.25,0.50,0.75$ from the average (Figure~\ref{fig:linedif}) 
highlights the presence of a narrow emission component.

To fit a detailed model of both disk and narrow line emission, we fixed 
the centroids of disk emission to those 
expected based on the mass ratio and velocity curve of Steeghs \&\ Casares 
(2002). The disk line is modelled using the method presented in 
Boroson et al. (2007), based on Horne (1995), allowing for turbulence and 
Keplerian shear. We set the components of the Mach turbulence matrix 
M$_{RR}$ and M$_{\theta\theta}$ to 1, and M$_{R\theta}$ to 0.5. This 
results in line profiles that are not sharply double-peaked. Horne
(1995) mentions that the M$_{ZR}$ component of the turbulence matrix
produces effects that are hard to distinguish from the M$_{RR}$
component,
and that correlated Z and R motion is similar to a radial wind rising
above the disk. Thus the effects of the turbulence model we assume may
mimic the result of an accretion disk wind. The models of Chiang
(2001) show that a disk wind model can fit the UV line profiles of
Hercules~X-1, which do not show clear double-peaked structure.

The stellar line is assumed to be a Gaussian with the 
same width at each orbital phase. The width is left as a free parameter of 
the fit. Although the Gaussian shape is not motivated by physical
considerations, it is a useful parameterization that allows us to
check our assumption of an emission component arising on the normal
star, and to obtain a crude measure of the width, flux, and
variability in such a component. In an earlier study of Hercules~X-1
(Boroson et al. 2000) we fit both the disk and donor star components
using Gaussians.
The continuum at 
each phase is scaled from the model we prepared before obtaining the 
observations. The normalizations of the disk and stellar lines are allowed 
to vary at each orbital phase.

We use a Nelder \&\ Mead (1965) downhill simplex method to fit the 
spectra in the neighborhood of the bright emission lines.
The best-fit parameters obtained by this method are shown in Table~3. For 
each line, we present fits obtained with the mass accretion rate fixed to 
give the Eddington luminosity for a 1.4~M$_\odot$ neutron star, 
L$_x=1.2\times10^{-8}$~M$_\odot$~yr$^{-1}$, and fits with $\dot{M}$ as a 
free parameter. The fits depend strongly on the inclination $i$ and
the radius of the disk edge $r_{\rm disk}$ combined in the disk edge
velocity $v_{\rm edge}=(G M_{\rm ns} / r_{\rm disk})^{1/2} \sin i$,
and only weakly on the inclination $i$ itself. In our fits, we have frozen the
inclination to $i=44^\circ$ as found for the inclination of the radio
jet by Fomalont et al. (2001), and present only $v_{\rm edge}$ in Table~3.

For the brightest emission line feature, the O\,VI doublet, 
the ratio of the narrow emission features (1032\AA\ line to 1038\AA\ 
line) is allowed to vary, while the ratio for the disk lines is determined 
from the disk models. The interstellar H$_2$ absorption is determined from 
fits to other spectral ranges, while interstellar atomic absorption from 
Si\,II$\lambda1020.70$, O\,I$\lambda1026.47$, C\,II$\lambda 1036.34$ and 
O\,I$\lambda1039.23$ are fit along with the emission lines, though they 
are constrained to be constant over time. The saturated H~Ly$\beta$ line 
at 1026.7\AA\  is also fit along with the emission lines. 

We plot the results in Figure~\ref{fig:plotlines}. The formal $\chi^2$ for 
the fit is $\chi^2=1.030\times10^{4}$ with $8896$ degrees of freedom, or a 
reduced $\chi^2$ of $\chi^2_\nu=1.16$. The narrow Gaussian lines from the 
normal star appear to be much weaker near $\phi=0$, so that in spite of 
the low inclination of the system, the heated spot on the donor star may 
be obscured. 

The best-fit projected velocity at the outer edge of the accretion disk, $v_{\rm 
edge}=210$~km~s$^{-1}$, implies a disk radius $r_{\rm disk}=1.9\times10^{11}$~cm for 
$i=44^\circ$. While significantly smaller than the result of 2.5$\times10^{11}$~cm found 
from fits to C\,IV and N\,V lines by Kallman et al. (1998), this is still unphysical as it 
is comparable to the distance of the L1 point from the neutron star, $\approx 
2\times10^{11}$~cm. The average radius of the Roche lobe of the neutron star should be 
1.5$\times10^{11}$~cm. If the disk inclination is 38$^\circ$, or 1$\sigma$ below the 
measured value of the jet inclination as favored for the orbital inclination by Steeghs 
\&\ Casares (2002) from orbital constraints, then the outer disk radius implied by the 
best-fit $v_{\rm edge}$ is 1.5$\times10^{11}$~cm.

In Figure~8 we show the total flux in the disk and narrow lines as a 
function of orbital phase and in Figure~9 we show the velocities of the 
Doppler shifts of the centroids of the narrow emission components of O\,IV 
as a function of orbital phase.

The results of the fit to O\,VI should be treated with caution because of 
the extensive modeling of interstellar lines required to match the data. 
The choice of a Gaussian shape for the stellar component probably does
not have much of an effect on our conclusions about the accretion
disk. When we use a Lorentzian shape instead, the average disk flux
only differs by 3\%\ and the disk flux at the 15 phases observed, as
determined by both fits in which the stellar component has a Gaussian
or Lorentzian shape, have a cross-correlation coefficient of 0.93. If
instead we fit our model to only half of the observations, we still
obtain a disk edge velocity of 200~km~s$^{-1}$ and a radial emissivity
power law constant of -0.98. Thus our fitted parameter values appear
not very sensitive either to our choice of the emission profile from
the donor star or from our sampling of the data. Comparing the
parameters obtained when the mass-accretion rate parameter is fixed to
the value giving the Eddington luminosity also gives a sense of how
sensitive our fits are to assumptions made about the system.

We test the model by applying it to the C\,III lines at 1175\AA. While 
there are no clear interstellar absorption features near 1175\AA, either 
atomic or molecular, the interaction between the 6 components of the lines 
may present complications. We calculate the line profiles using the 
oscillator strengths presented in Sokolov (2011), based on the 
Vienna Atomic Line Database (VALD-D, Kupka et al. 1999). A first 
fit, assuming 
the narrow components were optically thin, predicted that all emission was 
from the disk and none from the heated stellar surface, and the outer disk 
radius was twice as large (2.2$\times10^{11}$~cm) than the fits to O\,VI 
suggested. Because of the effective broadening of the multiplet components 
in C\,III, it is more difficult for narrow stellar emission features to 
stand out against the disk line, and a larger disk size (slower Keplerian 
velocities) could compensate for neglected stellar components.

To test whether the narrow stellar components suggested by the O\,VI fits 
could also be accomodated by the C\,III lines, we fixed the ratio of 
narrow stellar to broad disk emission at each phase to that found for 
O\,VI (the fit with the mass accretion rate through the disk as a free
parameter). This fit did not improve on the fit with all emission from the 
disk. At some phases, the narrow features would stick above the disk line 
and be noticeable. However, if we dropped the assumption that the narrow 
C\,III lines were optically thin, we obtained good fits, with a reduced 
$\chi^2=1.10$ and a disk edge velocity of $v_{rm edge}=190$~km~s$^{-1}$.

We fit the model to the Si\,IV doublet at 1393 and 1403\AA\ and 
the O\,V line at 1371\AA\ observed in 2010 with the HST COS. The lines 
from Sco~X-1 are subject to interstellar absorption from Ni\,II 
at 1370.13\AA, and Si\,IV itself. The model fits the spectra well 
(Figures~\ref{fig:silinefit} and \ref{fig:olinefit}), with similar values 
for disk radius, narrow line width, and narrow line velocity. The O\,V 
line shows a clearer double-peak structure than the Si\,IV doublet 
components. 
In contrast to the other lines, O\,V at 1371\AA\ is not a
resonance line; the lower level of the transition is an excited level.
As such, O\,V is less likely to absorb or scatter radiation. This may
explain why the disk line is more clearly double-peaked and why the
O\,V stellar line is more narrow than for the other transitions. 
We obtained a better fit for O\,V at 1371\AA\ by allowing as free 
parameters Mach turbulence matrix elements M$_{RR}$ and M$_{\theta\theta}$ 
and setting M$_{R\theta}=0$.

We also fit the N\,V doublet at 1239,1243\AA\ and the C\,IV doublet at 
1549,1551\AA\
observed with the HST GHRS, and confirm the result of Kallman, Boroson,
\&\ Vrtilek
(1998) that disk models assuming standard parameters do not fit well.
Our models match the spectra best for radial emissivity exponents 
$\gamma\sim0$ and narrow line components with Gaussian width far greater 
than in the fits to FUSE or COS spectra ($\sim1$\AA\ instead of 
$0.2-0.3$\AA). 

We include estimates of total line flux in Table~4, but we 
do not include the unphysical parameters of the fits to these lines in 
Table~3.
The fluxes in Table~4 have been corrected for 
overlapping interstellar absorption lines but should not otherwise depend 
strongly on our models for the emission lines.

\section{Bowen fluorescence and N\,III$\lambda991$ emission}

The Bowen Fluorescence process (Schachter, Filippenko, \&\ Kahn 1989) 
arises because of 
the nearly perfect coincidence of the \ion{He}{2} Ly$\alpha$ and the 
\ion{O}{3} 2p$^2$ -- 2p3d resonance line ($\lambda$304), resulting in 
\ion{O}{3} near--UV primary cascades at $\lambda\lambda$3133, 3444 and 
secondary cascades at $\lambda$374 back to the ground state.  If 
conditions are right, an additional fluorescence occurs, since the 
\ion{O}{3}$\lambda$374 line is almost coincident with the two \ion{N}{3} 
2p -- 3d resonance lines, resulting in N\,III optical primary cascades at 
$\lambda$4634, 4641, 4642.  The emission in all of these cascades is 
completely dominated by the Bowen process; detection of any of them is a 
clear confirmation.

In Boroson et al. (2007), we observed the N\,III$\lambda$990 line, an 
analogous ground-state Bowen cascade produced as a result of the 
\ion{N}{3} primary cascades. This detection confirmed the presence of the 
Bowen process and set limits on the size of the emitting region.

The intensity of N\,III$\lambda990$ resulting from Bowen fluorescence is 
related to the intensity of the Bowen lines at 4640~\AA by
\begin{equation}
I_f(991)=(4640/991) B(\mbox{3p2p}) I(4640)
\end{equation}
where $B(\mbox{3p2p})$ is the branching ratio of 2p-2s$^2$p$^2$ 2D versus
3p-3s. From CHIANTI, we take $B(\mbox{3p2p})=0.55$ and 
$B(\mbox{3d3p})=0.0058$.
The remainder of the line intensity is therefore from collisional 
excitation and is given by $I_c(991)=I(991)-I_f(991)$.
Our measurement of the intrinsic 991\AA\ line flux depends on the
E(B-V) reddening value we adopt, as well as our reconstruction of
the interstellar molecular and atomic absorption lines that overlap
with the 991\AA line. We take I(4640) from Schachter, Filippenko, \&\ Kahn 
(1989), 
although the observations were not contemporaneous.
From our estimate that without 
absorption lines the 
N\,III line flux would be $\sim5\times10^{-13}$~erg~cm$^{-1}$~s$^{-1}$, we 
find that the Bowen process would have contributed 17\%\ of this amount. 
For E(B-V)$=0.1,0.2,0.3,0.4$, the Bowen contributions to the dereddened 
line would be 4\%, 0.8\%, 0.2\%, and 0.04\%, respectively.

Another consequence of the N\,III cascade should be a multiplet near 
1184\AA.  The 2p3 2P states that feed the 2s2p2 2D
state also decay to the 2s2p2 2S and 2D states, resulting in unobservable 
lines at 772.9\AA\ and 1006\AA. Lines at 1182.97,1183.03,1184.51, and 
1184.57\AA, with relative strengths of 0.21, 0.33, 0.42, and 0.16, 
respectively, might be observed at the long-wavelength edge of the FUSE 
detectors. The photon flux in the 1184\AA\ multiplet should be 
67\%\ of the contribution to the 991\AA\ line. 
As shown in 
Figure~1, there is no hint of an upturn in flux near 1184\AA.

In Boroson et al. (2007), we also considered the ratio of N\,III at 
991\AA\ to C\,III at 977\AA\ as a diagnostic of the Bowen process 
contribution to the 991\AA\ line. For Sco~X-1, this is rendered difficult 
by the larger interstellar absorbing column. The C\,III line overlaps both 
a saturated absorption line from interstellar C\,III and also an O\,I line 
at 976.448\AA. Raymond (1993) found ratios between the N\,III line excited 
collisionally and C\,III$\lambda977$ of $I(991)/I(977)=0.34$ assuming 
cosmic abundance. If instead abundances enhanced by the CNO process were 
assumed, the ratio would be $I(991)/I(977)=0.53$. For Her~X-1 the ratio of 
$0.30\pm0.05$ did not require enhancement by Bowen fluorescence. In the 
case of Sco~X-1, our attempts to fit the C\,III line suggest 
$I(991)/I(977)\gae1$.

Given that even for E(B-V)=0, we expect
from non-contemporaneous observations of N\,III lines at 4640\AA\  that
only 17\%\ of the line flux arises from the Bowen process, and that the 
fluorescence lines at 1184\AA\ were not observed, it seems likely that
Bowen fluorescence cannot explain the large $I(991)/I(977)$ ratio. 
Instead, we suggest either that CNO processing has enhanced the N/C ratio 
beyond that assumed by Raymond (1993), or that our models have 
underestimated the C\,III flux at 977\AA, which may be concentrated at 
wavelengths that overlap with saturated interstellar absorption.
This conclusion depends on the suitability of the non-contemporaneous
observations for comparison. The optical lines at 4640\AA\ and
UV lines at 991\AA\ may be highly variable, and the values we have
adopted could have been measured when the source was in extreme states of line
flux.

\section{Discussion}

UV emission line variability, whether investigated through differences 
from the mean spectrum, through Doppler tomography, or through detailed 
line models, suggests that a portion of the emission arises from the 
accretion disk and that in addition, narrower emission lines are present. 
For much of the binary orbit, the narrow lines appear to move as expected 
from the X-ray heated face of the normal star. The spectra and models are 
not sufficient to determine whether deviations are the result of 
scattering of the resonance lines, perhaps in an outflowing wind, or 
whether another source of emission, for example a hot spot where the gas 
stream strikes the disk, may contribute.

When we compare the line fluxes with those predicted by Raymond (1993), 
as shown in Table~4, we 
find the rms difference is at a minimum for E(B-V)$=0$, but that the 
Pearson correlation is highest (for either the COS model with cosmic 
abundances or the CNO enhanced model) for E(B-V)$=0.2$, with the CNO model 
providing a higher correlation to the observed line fluxes (observed at 3 
separate times with 3 different instruments) than the COS model. For 
E(B-V)$=0.2$, the dereddened line fluxes observed are $\sim10\times$ 
greater than predicted by Raymond (1993). Some of this may be explained by 
the larger disk radius that we find.

We also note that the emission lines were not observed contemporaneously. 
Table~4 presents line fluxes measured at separate times using FUSE 
(N\,III, O\,VI, and C\,III), GHRS (N\,V and C\,IV), and COS (Si\,IV and 
O\,V). The Si\,IV line flux in the COS model is greater than the flux 
observed for E(B-V)=0.1, whereas the O\,VI flux is under-predicted. If the 
mass accretion rate were higher during the FUSE observation than the COS 
observation, this could explain both the higher O\,VI flux compared with 
the model and the greater disk radius.

The interpretation of the lines depends not only on the uncertain 
mass accretion rate and amount 
of interstellar extinction (we show fluxes dereddened for 
E(B-V)$=0.1,0.2,0.3,0.4$) but also on the details of the reddening curve 
in the far UV range. We have assumed the CCM reddening curve (Cardelli, 
Clayton, \&\ Mathis, 1989), scaling between E(B-V) and A(V) by using 
$R=A(V)/E(B-V)=3.1$, but there the 
column towards Sco~X-1 may have unusual reddening in the far UV, similar 
to $\sigma$~Sco (Schachter, Filippenko, \&\ Kahn 1989). We examined the 
reddening curves fit by Gordon, Cartledge, \&\ Clayton (2009) to 75 sight 
lines using FUSE spectra, and find in particular that the sample star 
closest to Sco~X-1, HD~147888, does not require a weaker correction for 
far UV wavelengths than typical, although the difference in the fit 
parameter that determines this effect, $C^{A(V)}_4$, is only 
$\approx1\sigma$. Still, we find that the details of the redening curve 
may contribute some systematic uncertainty to the line estimates beyond 
the uncertainty in the quantity of reddening. For example, for E(B-V)=0.1, 
using the CCM model, the dereddening of N\,III at 991\AA\ is 2.2 times 
greater than the 
dereddening of C\,IV at 1550\AA. Using instead the parameters for 
HD~147888, including $R=4.0$, the dereddening is only 1.8 times great for 
N\,III. 

Our disk line models very crudely estimate the mass flow through the 
accretion disk. The optical depths of the lines follow from the density 
determined by $\dot{M}$ and our assumed atomic abundance and ion fraction. 
We do not model the ionization in the disk, but assume the peak fraction 
given by Model~5 of the photoionization calculations of Kallman \&\ McCray 
(1982). The N\,III lines near 990\AA\ give a lower $\dot{M}$ estimate than 
the other lines, but this blend includes a non-resonance line.

The Bowen process does not appear to contribute to the N\,III line, either 
based on the strength of the feature expected from the optical lines or on 
the absence of features near 1184\AA. The strength of the N\,III line 
relative to the C\,III line at 977\AA\ provides more circumstantial 
evidence for CNO enhancement, although the 977\AA\ line, in common with 
much of the spectrum observed with FUSE, coincides with saturated 
instellar absorption features.

\section{Conclusions}

Although the far UV spectrum of Sco~X-1 is subject to many
interstellar molecular and atomic absorption features, the emission
lines may be resolved into components similar to those seen in
Hercules~X-1 (Boroson et al. 2007): narrow lines associated with the
X-ray illuminated face of the donor star and broad lines associated
with the accretion disk around the compact object. In further work
(Boroson et al. 2014, in preparation), we will present similar results
for the Z-source Cygnus~X-2, observed with the HST~COS.

The broad emission lines do not display a clear double-peaked
shape. Our models show that turbulence or a disk wind could blur out
these peaks. 

For some emission lines the fits are not good or
imply a disk larger than the Roche lobe of the neutron star.
The widths of the
broad lines are easier to accomodate if the inclination of the disk
is at the low end of the jet inclination range of 44$\pm6^\circ$. Among the
idealizations we have made in our disk model is the assumption that
the disk is flat and not warped. A warped disk, as generally accepted
in the Her~X-1 system as an explanation of its 35-day X-ray cycle,
would present different inclinations at different radii. A disk warp
may shield portions of the disk from direct X-ray illumination as
well. The current observations cannot constrain the full
three-dimensional variable shape of the disk to confirm models of a
disk warp.

We find no clear P~Cygni absorption feature from a disk wind or an 
X-ray induced wind
from the donor star, as we found
in Her~X-1 (Boroson et
al. 2001). Sco~X-1 has a lower inclination so such winds may not
cover the line of sight towards emitting regions.
Furthermore, X-rays from the neutron star may remove ions responsible
for the UV resonance lines outside of the shadow of the accretion
disk. The
low inclination of Sco~X-1 also limits how much of the accretion
disk shadow covers the emitting regions.

Much remains to be learned about accretion in Sco~X-1:
the extent and role of disk winds versus turbulence, why some lines
are fit poorly or require a disk too large, and the extent of 
disk warping, and what is changing as the X-rays move along the
Z~track. 
The low orbital inclination prevents the use of eclipse
mapping as a tool. Instead, the response of the UV emission lines to
the stochastic variability of the X-ray source may probe both the
X-ray variability (for example, how isotropic it is) and 
the varying shape and physical conditions of the UV reprocessing
regions. Future observations coordinating simultaneous X-ray and
UV spectroscopy are encouraged.

 \acknowledgements

Based on observations made with the NASA-CNES-CSA Far Ultraviolet 
Spectroscopic Explorer. FUSE is operated for NASA by the Johns Hopkins 
University under NASA contract NAS5-32985.
Based on observations with the NASA/ESA {\it Hubble Space Telescope}, 
obtained at the Space Telescope Science Institute, which is operated by 
the Association of Universities for Research in Astronomy, Inc., under 
NASA contract GO-05874.01-94A. We thank D. Steeghs for providing
optical spectra of Sco~X-1.

\clearpage

\begin{table}
\caption{FUSE Observation Log \label{tab:exposures}}
\begin{tabular}{llllcl}
Orbit & File Identifier & MJD (start) & MJD (end) & Duration (s) & Orbital Phase\\
\hline
      1 & B0060101001 & 52140.2496 &  52140.2669 &  1490.
 &       0.493\\
       2 & B0060101002 &  52140.3191 &  52140.3363 &  1485.
 &       0.582\\
       3 & B0060101003 & 52140.3882 &  52140.4057 &  1509.
 &       0.671\\
       4 & B0060101004 & 52140.4577 &  52140.4751 &  1503.
 &       0.761\\
       5 & B0060101005 & 52140.5273 &  52140.5444 &  1482.
 &       0.845\\
       6 & B0060101006 &  52140.5965 &  52140.6138 &  1496.
 &       0.934\\
       7 & B0060101007 & 52140.6658 &  52140.6832 &  1508.
 &      0.019\\
       8 & B0060101008 & 52140.7352 &  52140.7526 &  1506.
 &       0.108\\
       9 & B0060101009 & 52140.8046 &  52140.8220 &  1504.
 &       0.197\\
      10 & B0060101010 & 52140.8740 &  52140.8914 &  1505.
 &       0.287\\
      11 & B0060101011 & 52140.9433 &  52140.9608 &  1508.
 &       0.376\\
      12 & B0060101012 & 52141.0128 &  52141.0302 &  1501.
 &       0.465\\
      13 & B0060101013 & 52141.0822 &  52141.0996 &  1506.
 &       0.550\\
      14 & B0060101014 & 52141.1515 &  52141.1690 &  1513.
 &       0.639\\
      15 & B0060101015 & 52141.2209 &  52141.2372 &  1408.
 &       0.728\\

\end{tabular}
\end{table}

\begin{table}
\caption{HST COS Observation Log \label{tab:exposures2}}
\begin{tabular}{llllcl}
Orbit  & File Identifier & MJD (start) & MJD (end) & Duration (s) & Orbital Phase\\
\hline
      1 & lb2m01iwq & 55369.5198 &  55369.5247 &  422.
 &       0.122\\
      2 & lb2m01iyq & 55369.5563 & 55369.5612 & 422. 
 &      0.166\\
      3 & lb2m01j0q & 55369.5630 & 55369.5678 & 422.
  &     0.171\\
      4 & lb2m01j2q & 55369.5697 & 55369.5746 & 422.
  &     0.181\\
\end{tabular}
\end{table}

\begin{deluxetable}{lrrrrrrrrr}
\rotate


\tablecaption{Parameters of Line Fits}

\tablenum{3}

\tablehead{\colhead{Ion} & \colhead{$\lambda$} & 
\colhead{$a$,$a_i$\tablenotemark{a}} & \colhead{v$_{\rm 
edge}$\tablenotemark{b}} 
 & \colhead{$\gamma$ \tablenotemark{c}} 
& 
\colhead{$\sigma$ \tablenotemark{d}} & \colhead{Narrow} & 
\colhead{Disk $\dot{M}$} & \colhead{$\chi^2_\nu$, $\nu$} \\ 
\colhead{} & \colhead{(\AA)} & \colhead{} & \colhead{(km/s)} 
 & \colhead{} & \colhead{(km~s$^{-1}$)} & 
\colhead{Fraction} & 
\colhead{(M$_\odot$ yr$^{-1}$)} & \colhead{} } 

\startdata
N\,III & 990 &9.1(-5),0.3 & 140 & -1.44 &  & 0.01 & 
1.2(-8) & 0.91,4129 
\\
N\,III & 990 & 9.1(-5),0.3 & 150 & -1.44 & & 0.01 & 
1.7(-9) & 0.91,4130\\
O\,VI & 1035 & 7(-4),0.1 & 220 & -1.07 & 90 & 0.11 & 1.2(-8) &
1.22,8897\\
O\,VI & 1035 & 7(-4),0.1 & 210 & -0.98 & 100 & 0.09 & 6.3(-8) & 
1.16,8896 \\
C\,III & 1175 & 3.3(-4),0.5 & 180 & -0.93 & 140 & 0.11 & 1.2(-8) 
&  1.10,4434\\
C\,III & 1175 & 3.3(-4),0.5 & 190 & -1.02 & 140 & 0.11 & 4.3(-9) 
& 1.10,4433 
\\
O\,V & 1371 & 7(-4),0.3 & 280 & -1.07 & 50 & 0.10 & 1.2(-8) & 1.14, 
8002 \\
O\,V & 1371 & 7(-4),0.3 & 330 & -1.08 & 50 & 0.10 & 1.6(-8) & 1.12, 
8001 \\
Si\,IV & 1400 & 3.3(-5),0.3 & 250 & -0.92 & 120 & 0.16 & 1.2(-8) 
& 
1.31,12413\\
Si\,IV & 1400 & 3.3(-5),0.3 & 240 & -0.82 & 120 & 0.12 & 4.5(-8) 
& 
1.29,12412\\
\enddata

\tablenotetext{a}{Assumed atomic abundance and ion fraction, respectively}
\tablenotetext{b}{The projected velocity at the outer edge of the 
accretion disk. Based on two parameters of the fit, the disk radius and 
inclination, which only have a weak effect on the line model 
individually.}
\tablenotetext{c}{Radial power law index of disk 
emissivity}
\tablenotetext{d}{Gaussian width of narrow line (not resolved for N\,III)}
\tablecomments{Lines with 
wavelengths $\lambda<1200$ were observed with FUSE and lines 
with $\lambda>1200$ were observed with the HST COS.}


\end{deluxetable}

\begin{deluxetable}{cccccccc}




\tablecaption{UV Line Fluxes and Reddening}

\tablenum{4}

\tablehead{\colhead{E(B-V)} & \colhead{N\,III} & \colhead{O\,VI} & 
\colhead{C\,III} & \colhead{N\,V} & \colhead{O\,V} & \colhead{Si\,IV} & 
\colhead{C\,IV} \\ 
\colhead{} & \colhead{(990\AA)} & \colhead{(1035\AA)} & 
\colhead{(1175\AA)} & \colhead{(1240\AA)} & \colhead{(1371\AA)} & 
\colhead{(1400\AA)} & \colhead{(1550\AA)} } 

\startdata
0.0 & 5.0(-13) & 3.4(-12) & 1.95(-12) & 4.5(-12) & 2.9(-13) & 5.6(-13) & 
4.4(-12) \\
0.1 & 2.3(-12) & 1.3(-11) & 5.6(-12) & 1.2(-11) & 6.7(-13) & 1.3(-12) & 
9.3(-12) \\
0.2 & 1.1(-11) & 5.3(-11) & 1.6(-11) & 3.1(-11) & 1.5(-12) & 2.8(-12) & 
1.9(-11) \\
0.3 & 5.1(-11) & 2.1(-10) & 4.7(-11) & 8.2(-11) & 3.5(-12) & 6.7(-12) & 
4.1(-11) \\
0.4 & 2.4(-10) & 8.3(-10) & 1.4(-10) & 2.1(-10) & 8.1(-12) & 1.4(-11) & 
8.6(-11) \\
COS\tablenotemark{a} & 2.9(-12) & 7.1(-12) & 3.8(-12) & 4.2(-12) & 
2.3(-13) & 2.0(-12) &
4.0(-12)\\
CNO & 1.6(-12) & 3.4(-12) & 1.8(-12) & 2.3(-12) & 1.7(-13) & 6.2(-13) &
2.0(-12)\\
\enddata

\tablenotetext{a}{COS and CNO are fluxes based on Raymond (1993) and a 
distance of 2.8 kpc to Sco~X-1, assuming cosmic and CNO processed 
abundances, respectively}

\tablecomments{Fluxes in the emission lines as 
observed by FUSE 
($\lambda<1200$), the HST COS (O\,V and Si\,IV) and the HST GHRS (N\,V and 
C\,IV), and dereddened based on the CCM model (Cardelli, Clayton, 
\&\ Mathis, 1989).}



\end{deluxetable}

\clearpage

\setcounter{figure}{0}


\begin{figure}
\epsscale{0.7}
\plotone{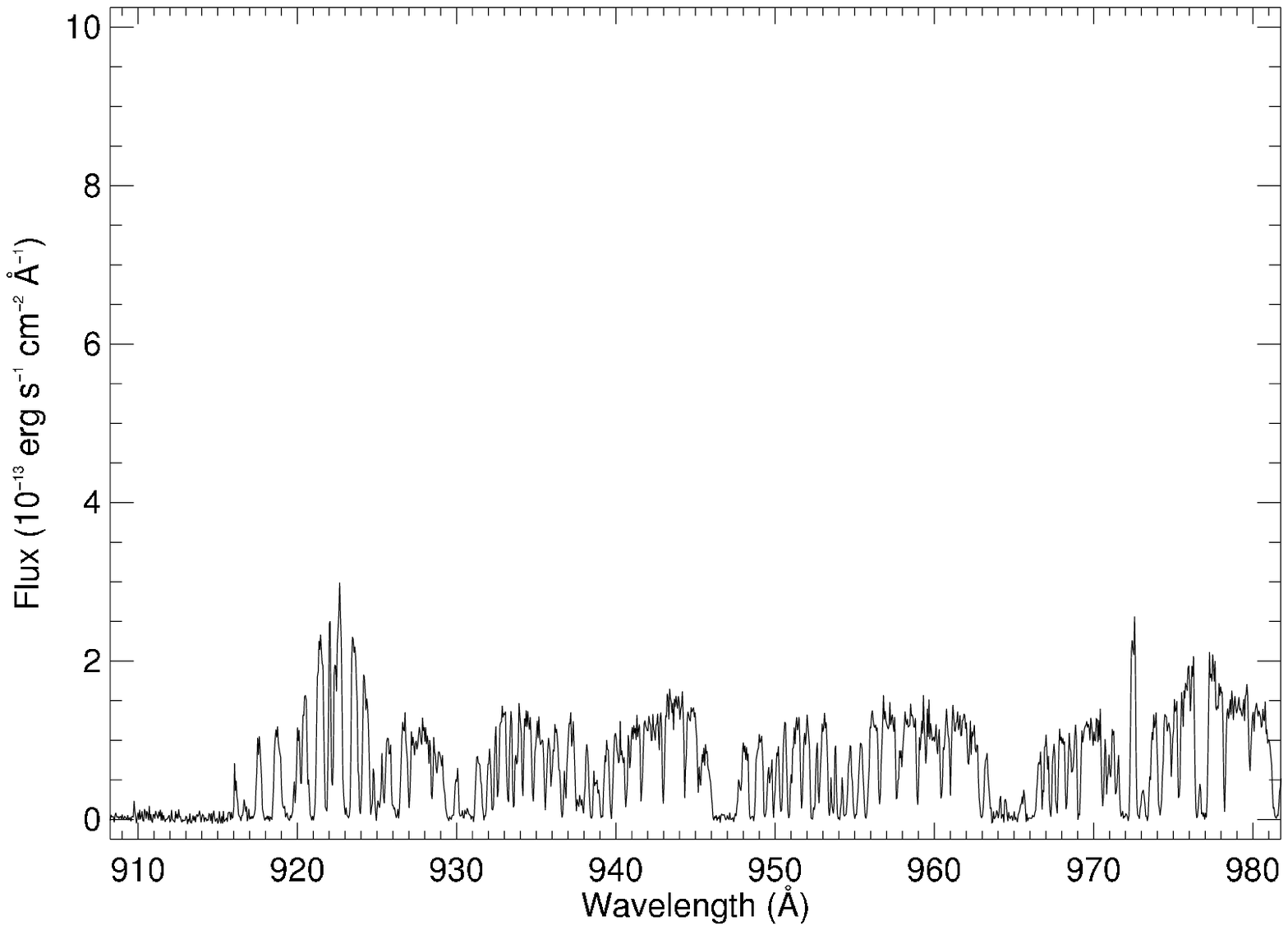}
\plotone{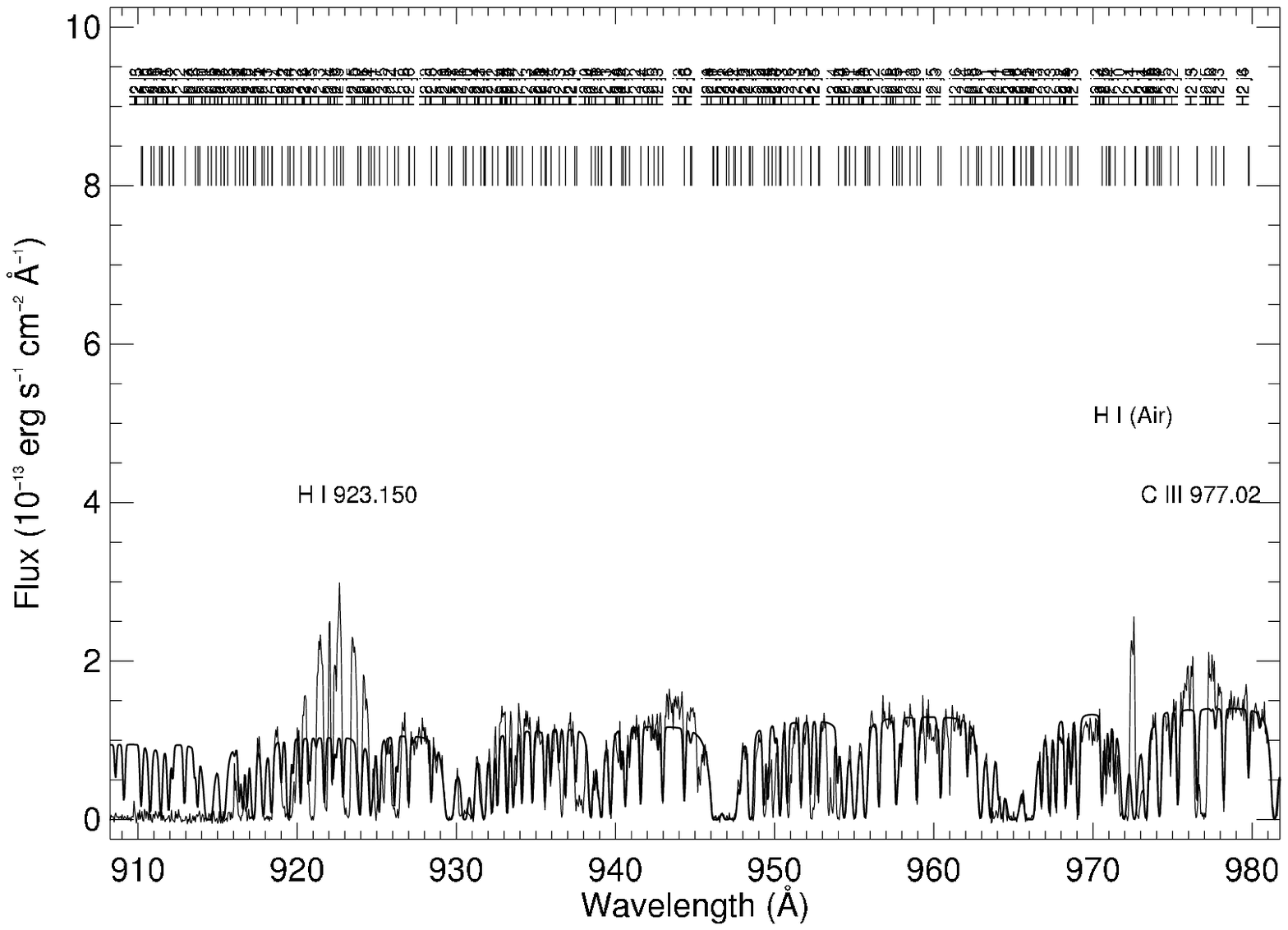}
\end{figure}

\newpage
\setcounter{figure}{0}
\begin{figure}
\epsscale{0.7}
\plotone{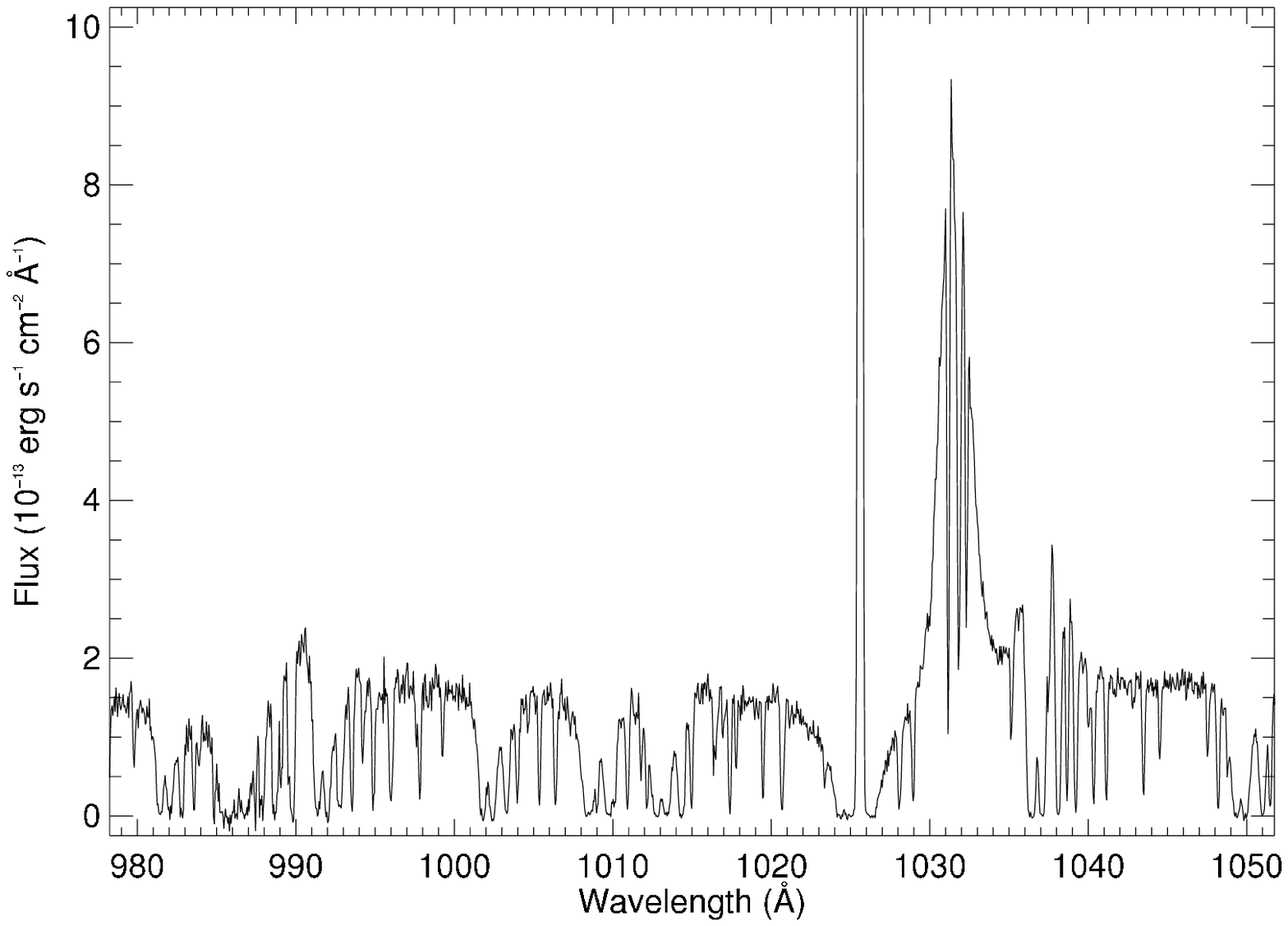}
\plotone{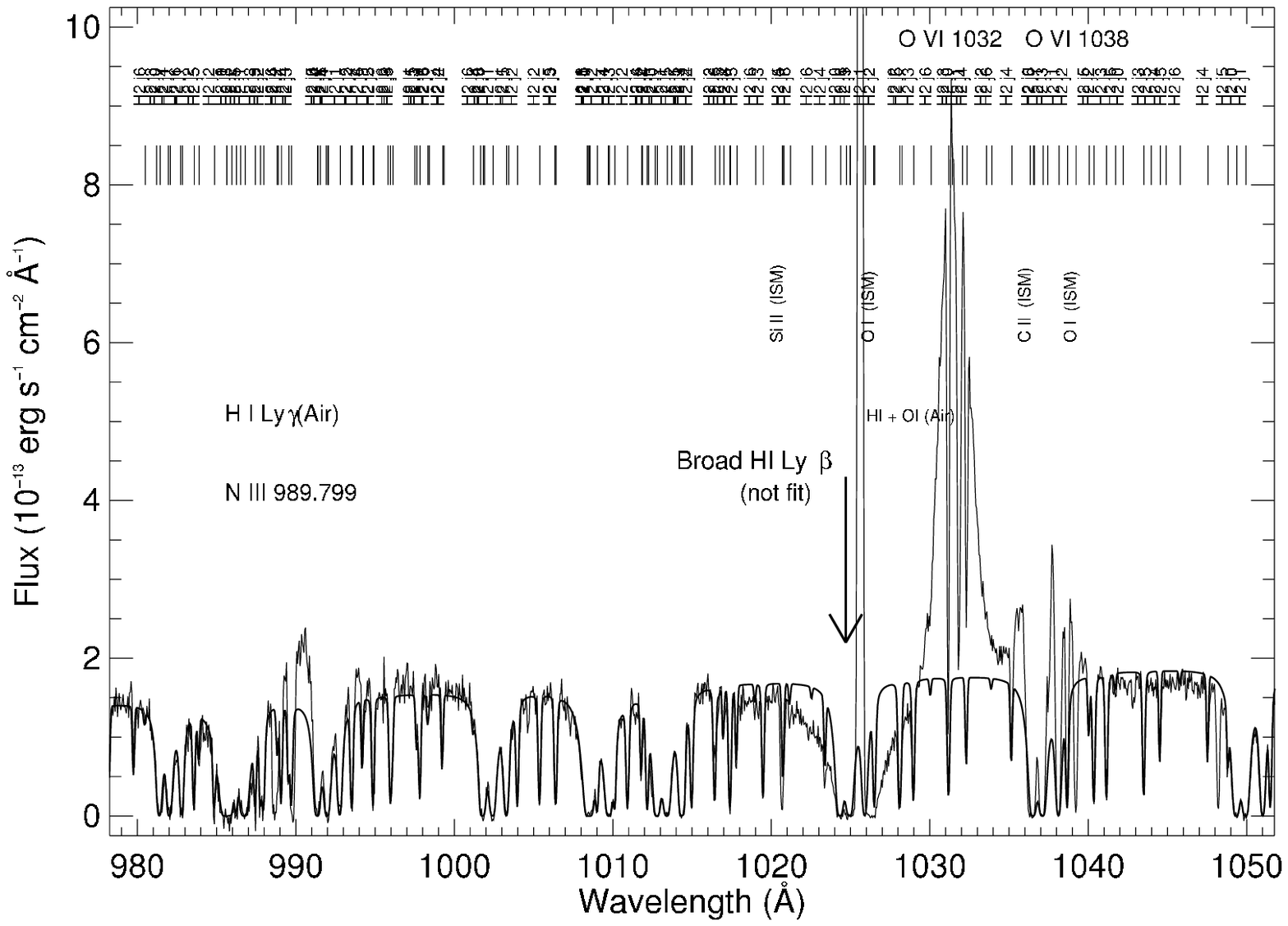}
\end{figure}

\newpage
\setcounter{figure}{0}
\begin{figure}
\epsscale{0.7}
\plotone{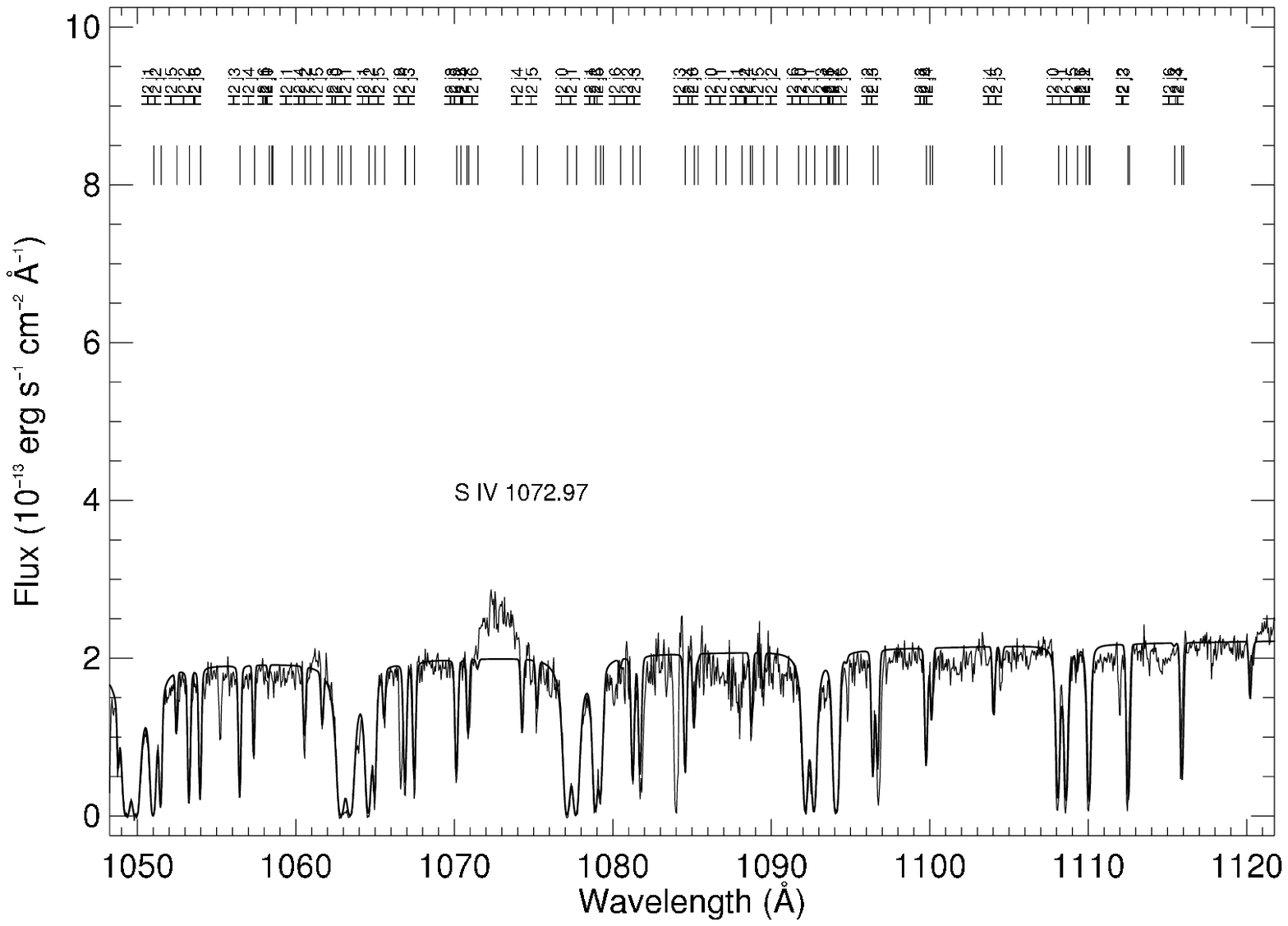}
\plotone{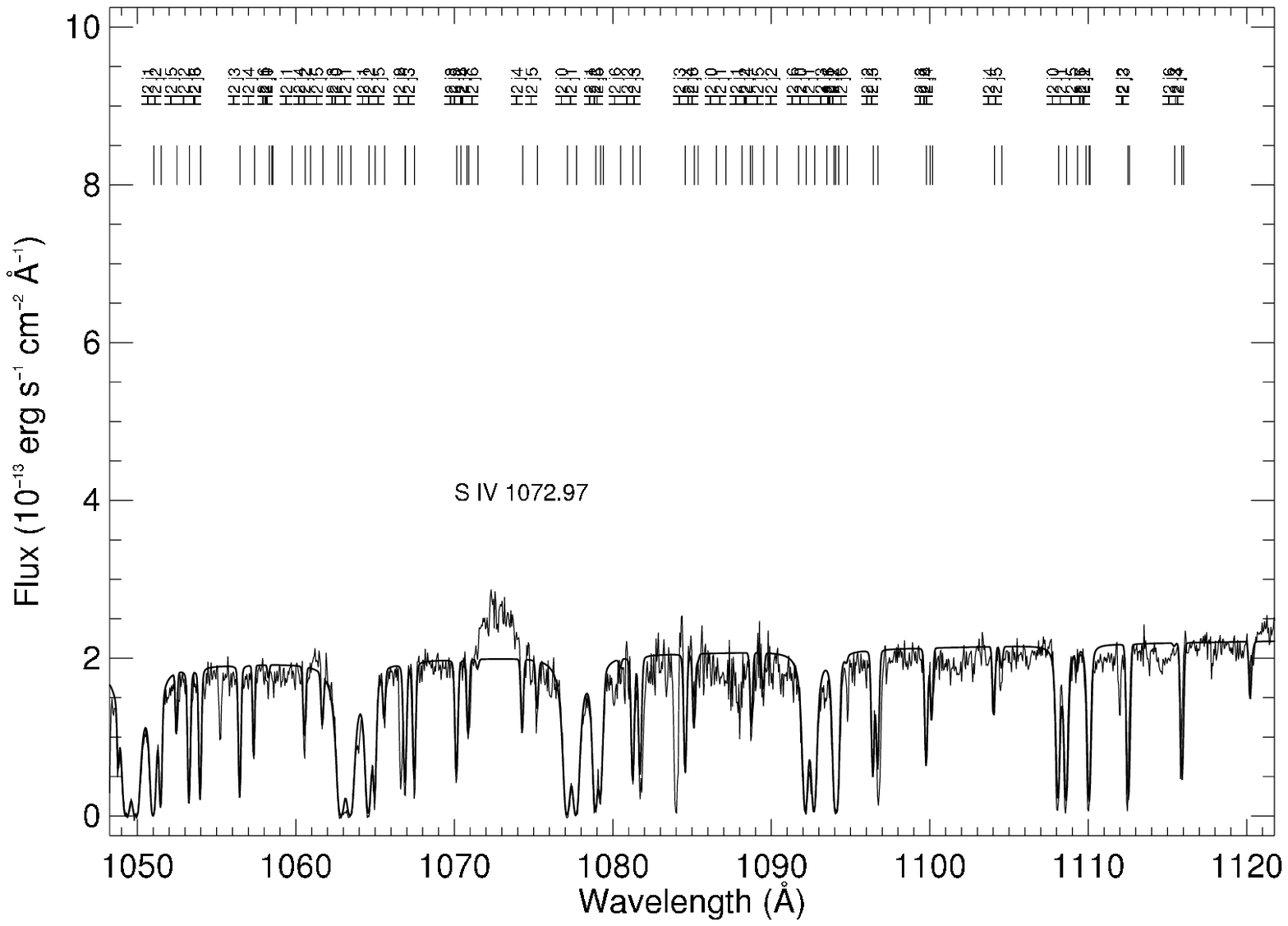}
\end{figure}

\newpage
\setcounter{figure}{0}
\begin{figure}
\epsscale{0.8}
\plotone{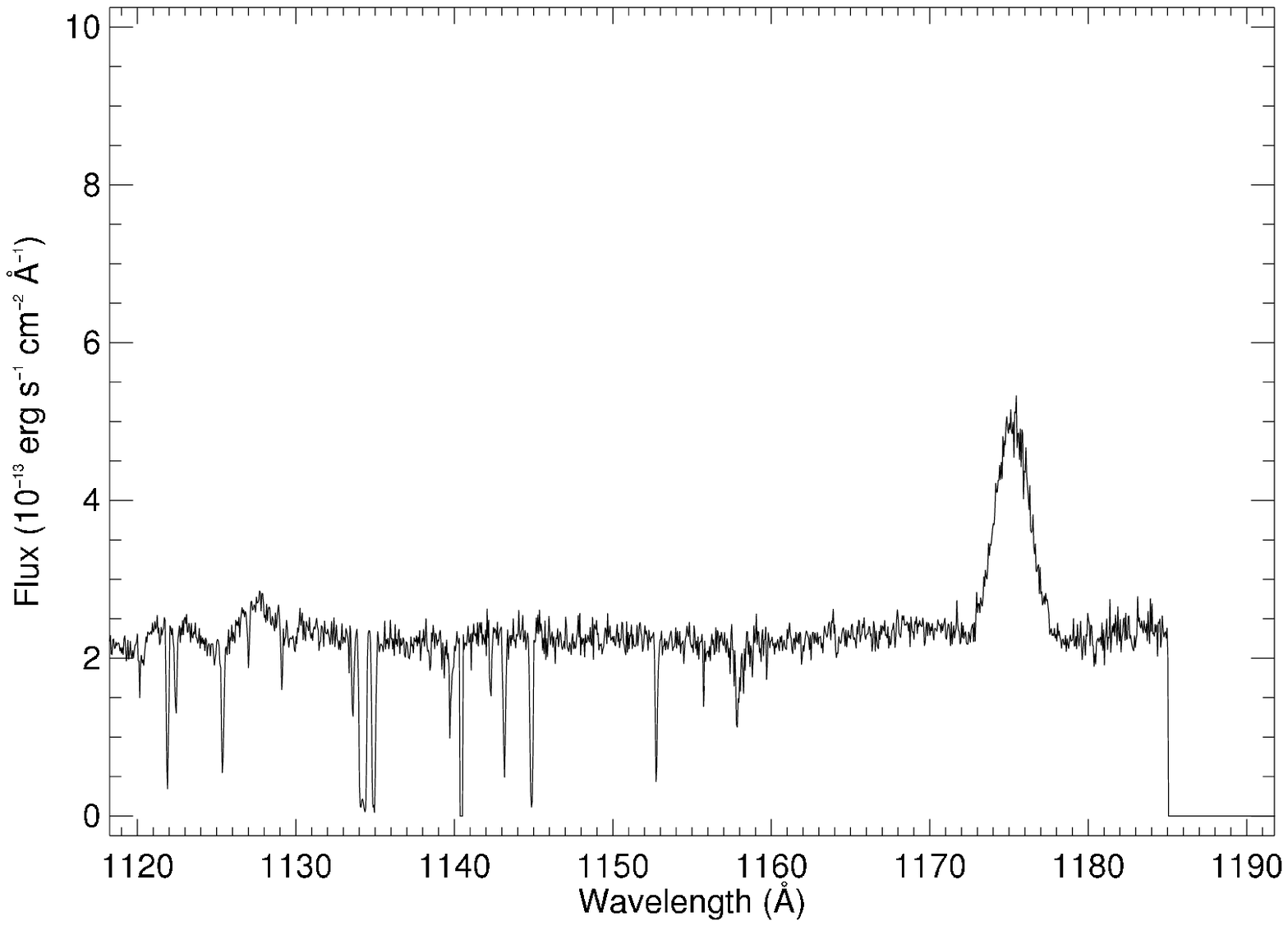}
\plotone{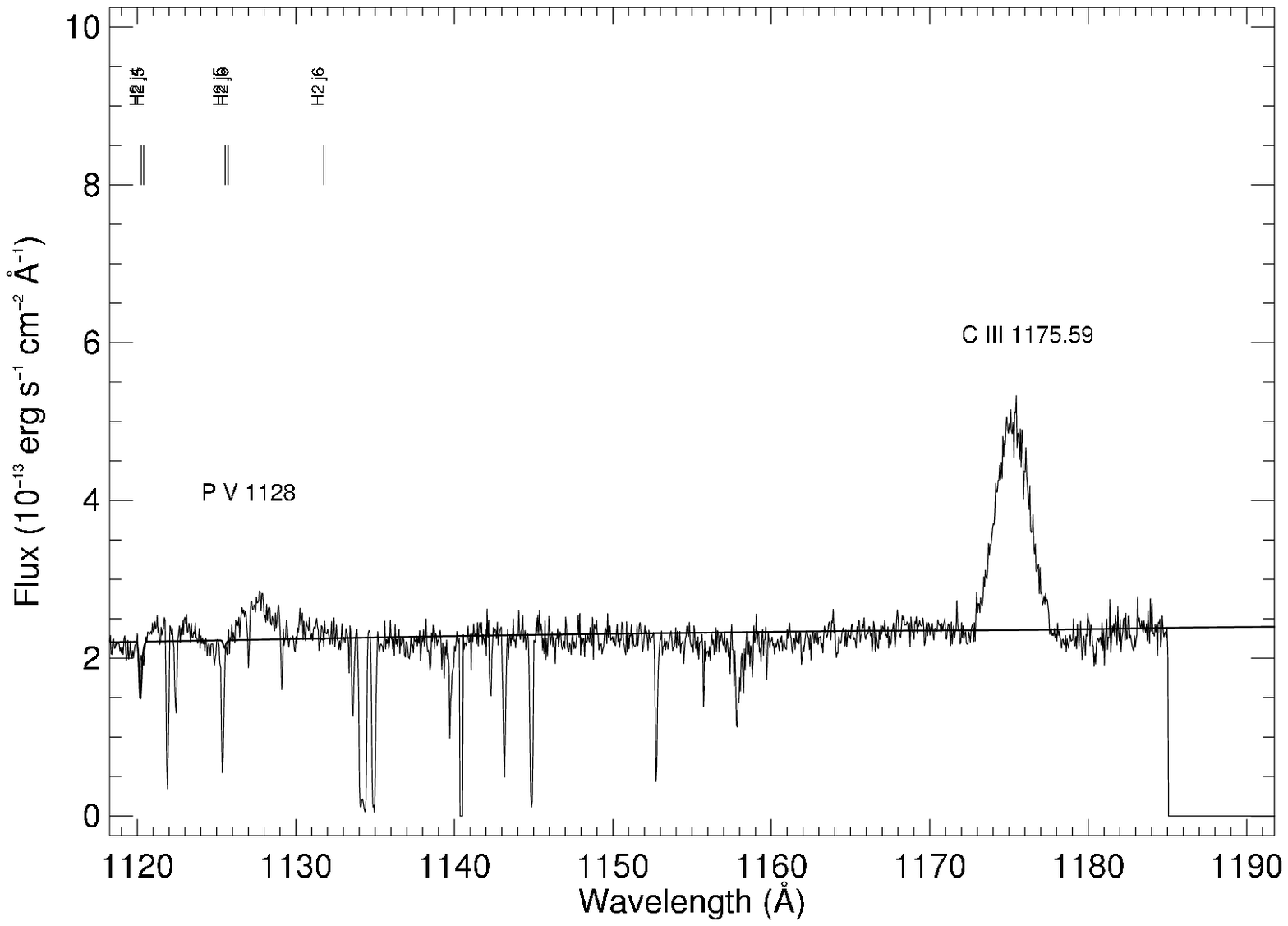}
\figcaption{In alternating panels, the average FUSE spectrum of 
Sco~X-1, and the spectrum with models of continuum and interstellar 
molecular absorption lines, together with identification of 
molecular and atomic 
interstellar lines and bright 
emission lines from the Sco~X-1 system. 
The broad
Lyman$\beta$ absorption line at 1026\AA\ is not fit here. In Figure~3, we 
fit this 
feature along with the variable O\,VI emission from the Sco~X-1 system, 
with which it overlaps.}
\end{figure}

\begin{figure}
\plotone{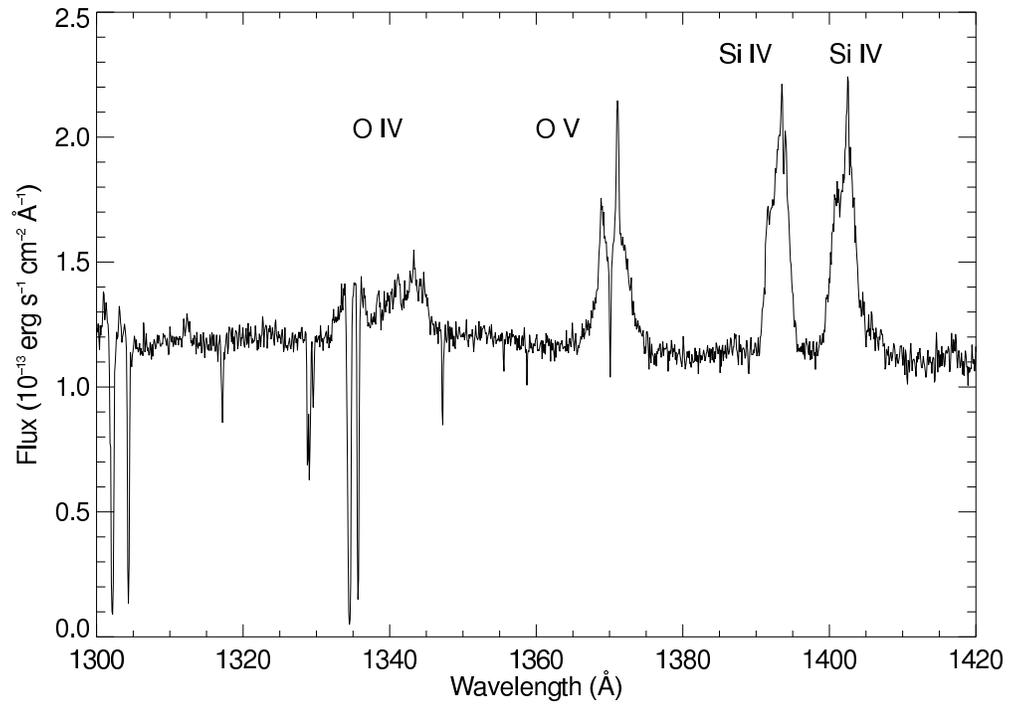}
\caption{The average COS spectrum of Sco X-1 from June 22, 2010, rebinned 
by a factor of 9 in wavelength. Strong emission lines are indicated.
\label{fig:cos}}
\end{figure}

\begin{figure}
\plottwo{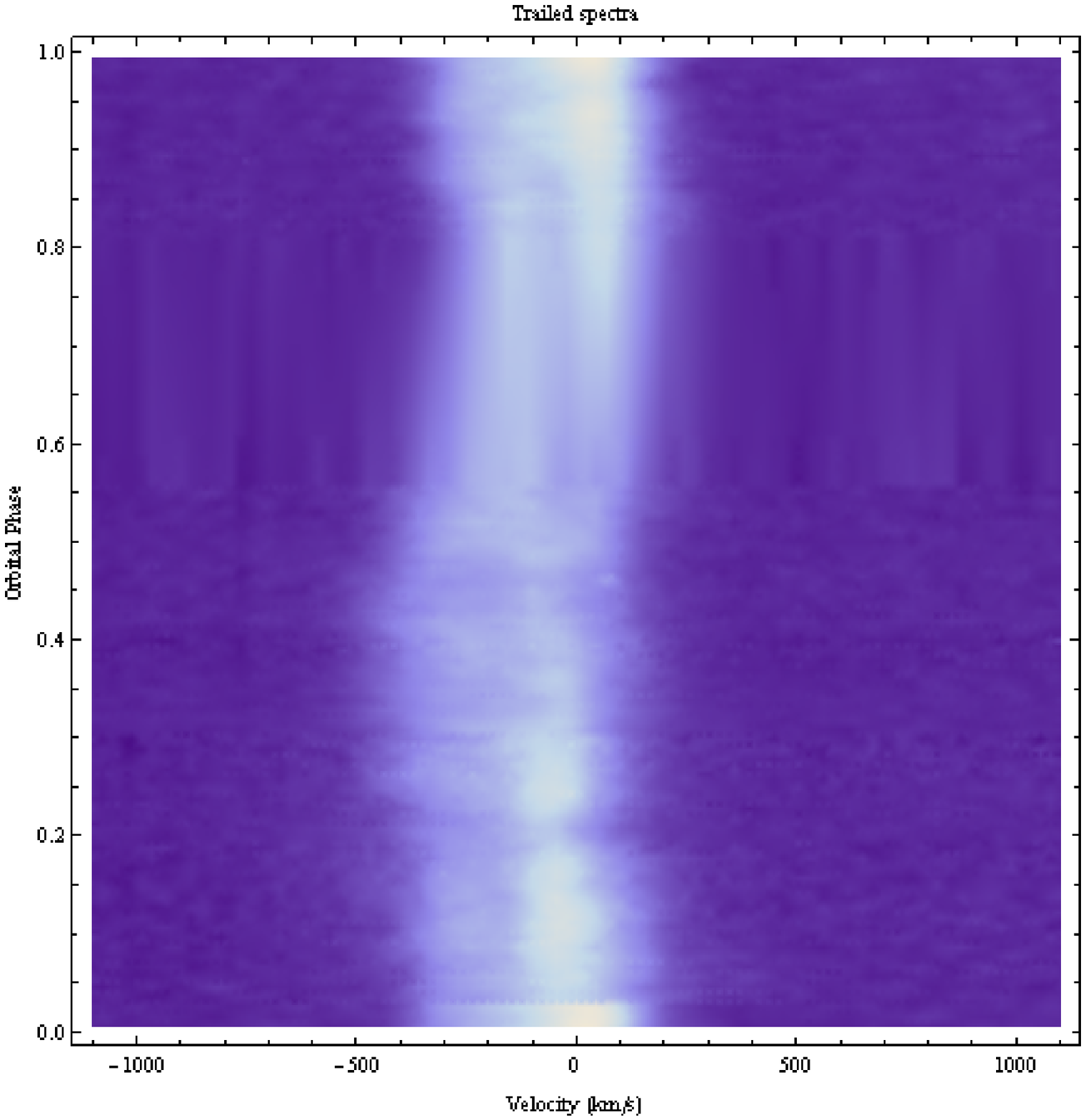}{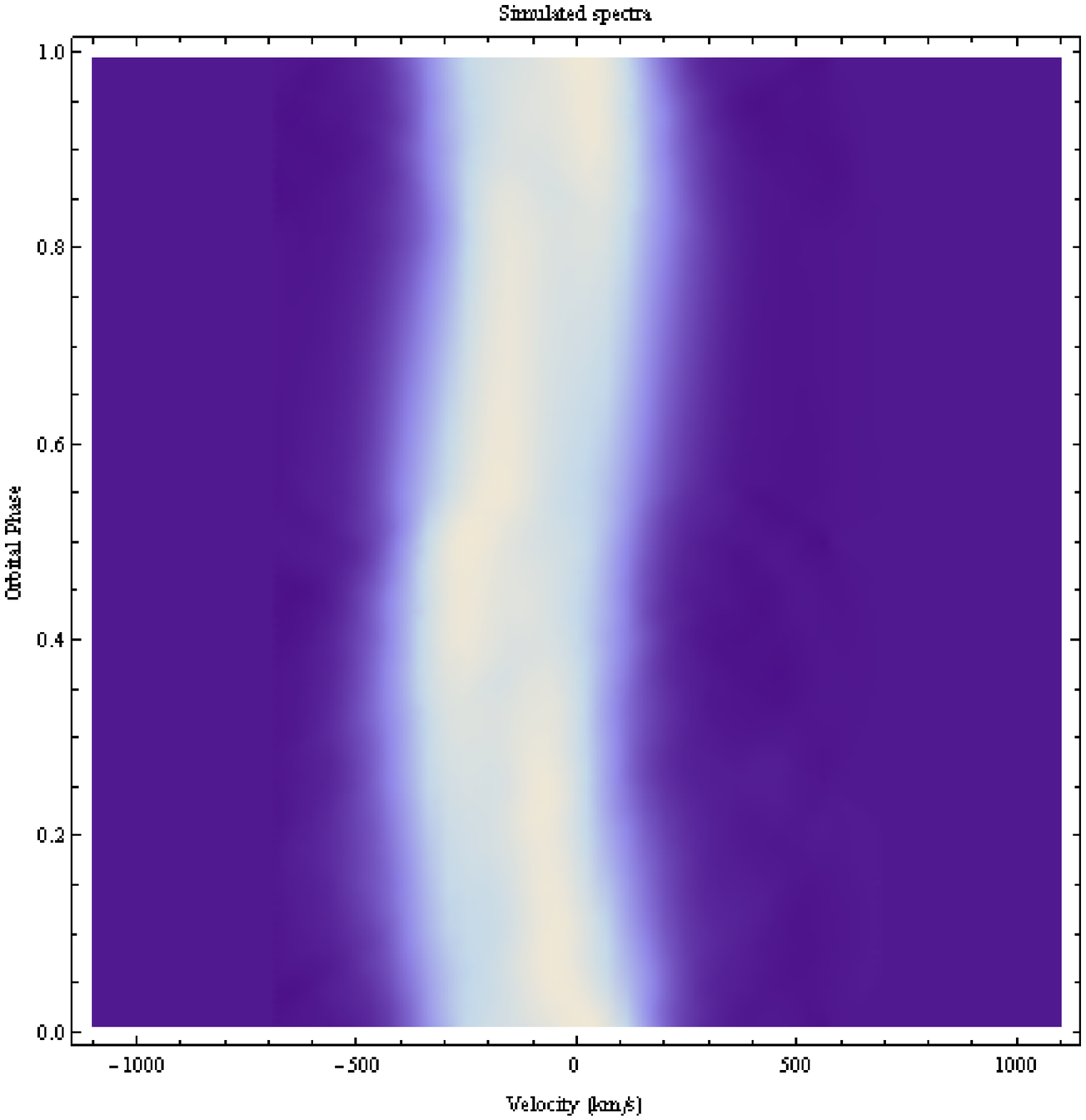}
\plotone{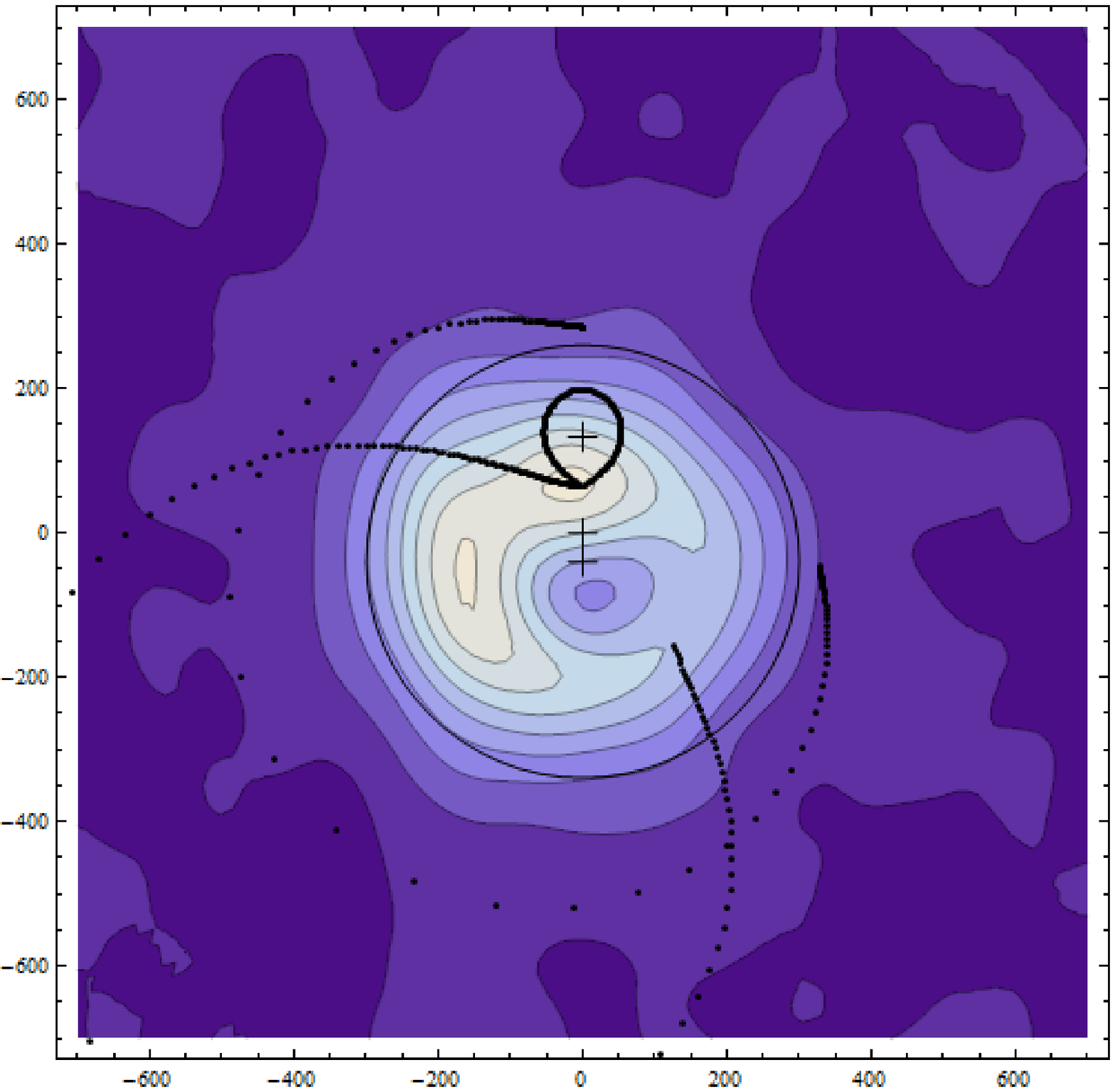}
\figcaption{Fourier-filtered doppler tomograms of the 
optical He\,II $4686$\AA\ 
line. The data from Steeghs \&\ Casares (2002) were
processed by the same methods that we used on the O\,VI line at 
1032\AA. We show velocities expected from the 
the donor star along with
the gas stream and the disk velocity along the gas stream.
The dots along the gas stream are spaced at $\approx13$~s intervals.
The center of mass of the stars are
indicated by ``+'' signs. The circle centered on the neutron star
shows the outer edge of an accretion disk of radius 10$^{11}$~cm.
We assume the centers of mass are
separated by
3.15$\times10^{11}$~cm, the mass ratio is $M_{\rm normal}/M_{\rm
ns}=0.42$, and the inclination is 44$^\circ$.
\label{opticaltomo}
}
\end{figure}

\begin{figure}
\plottwo{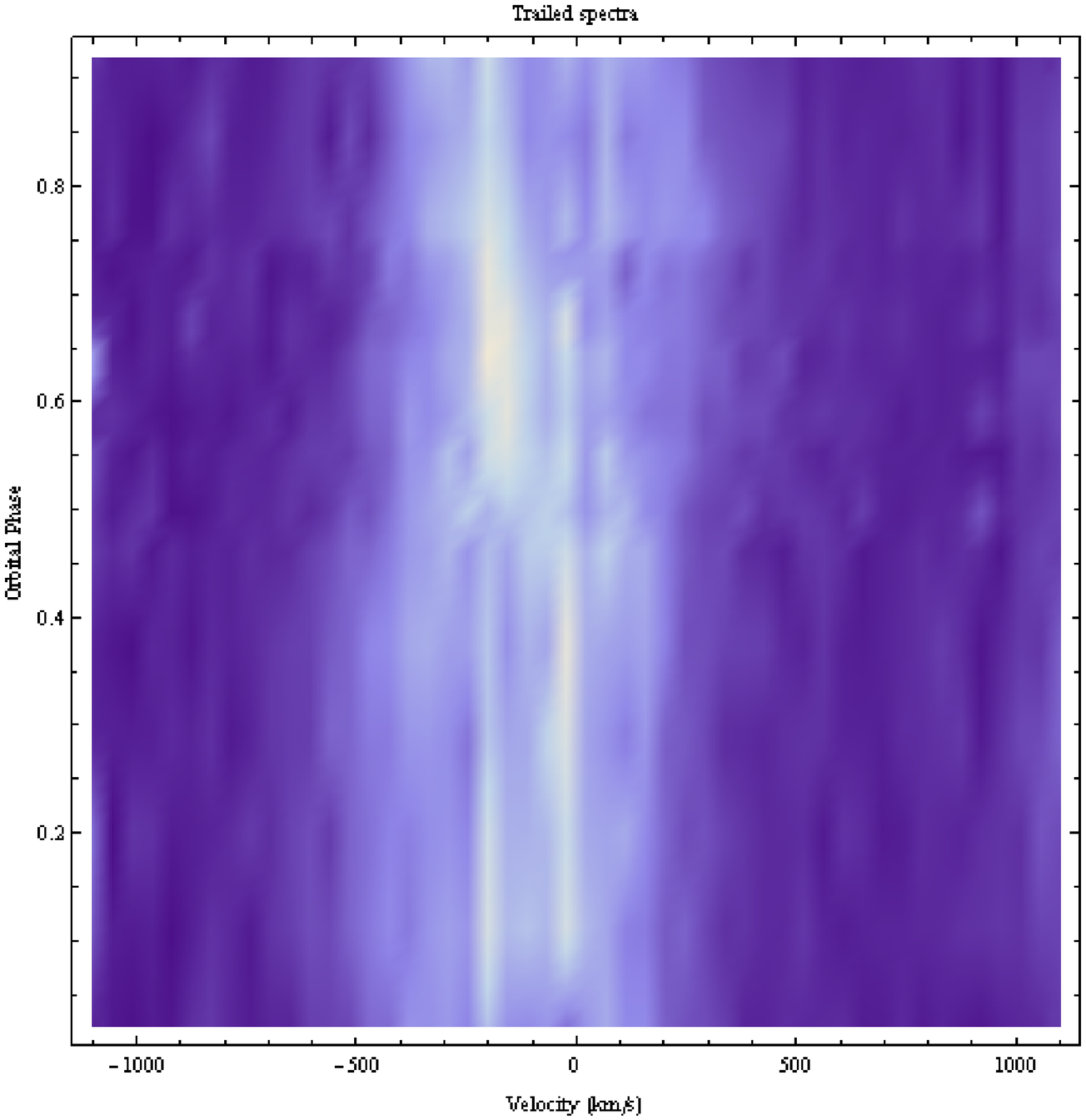}{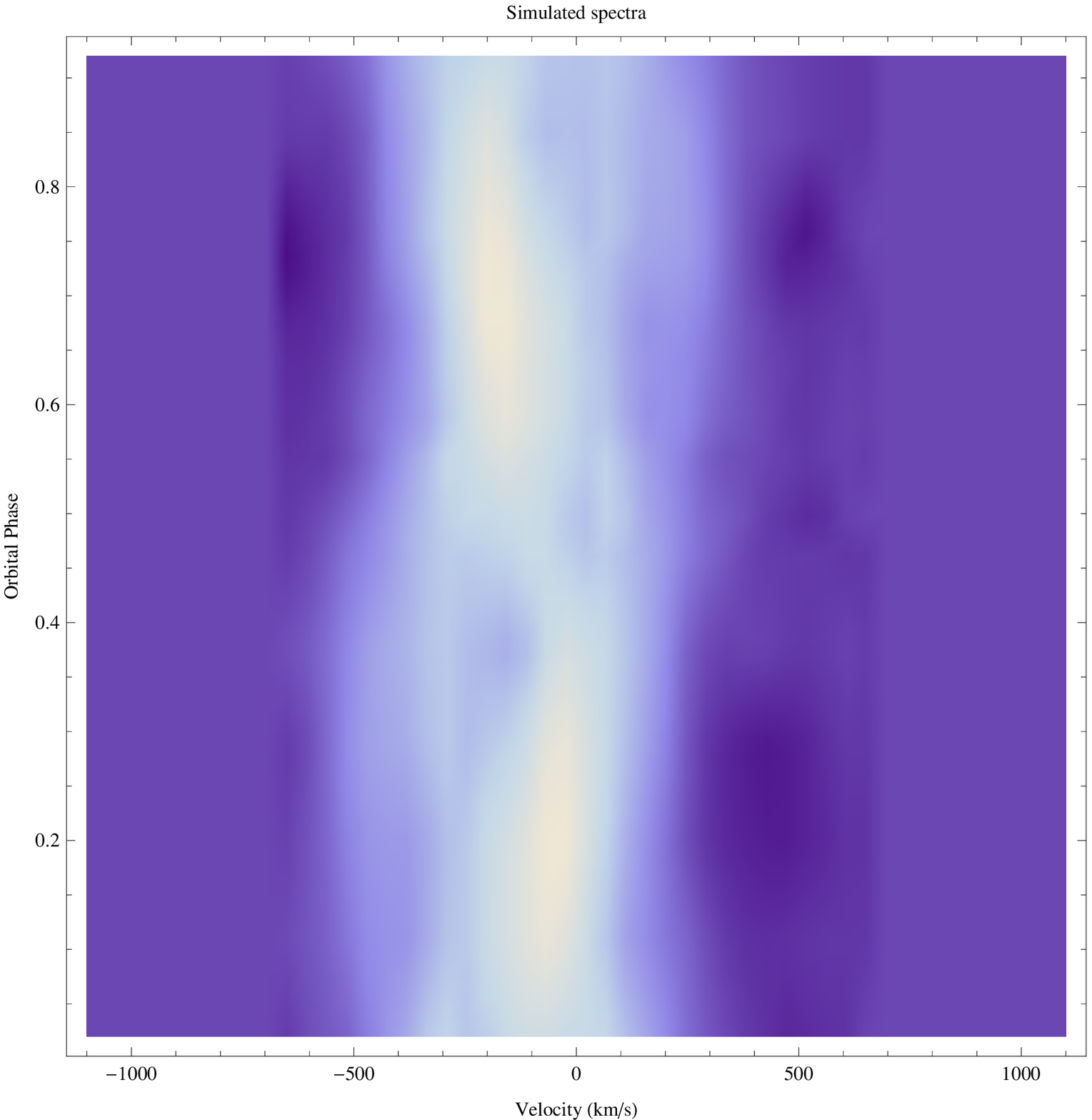}
\epsscale{0.8}
\plotone{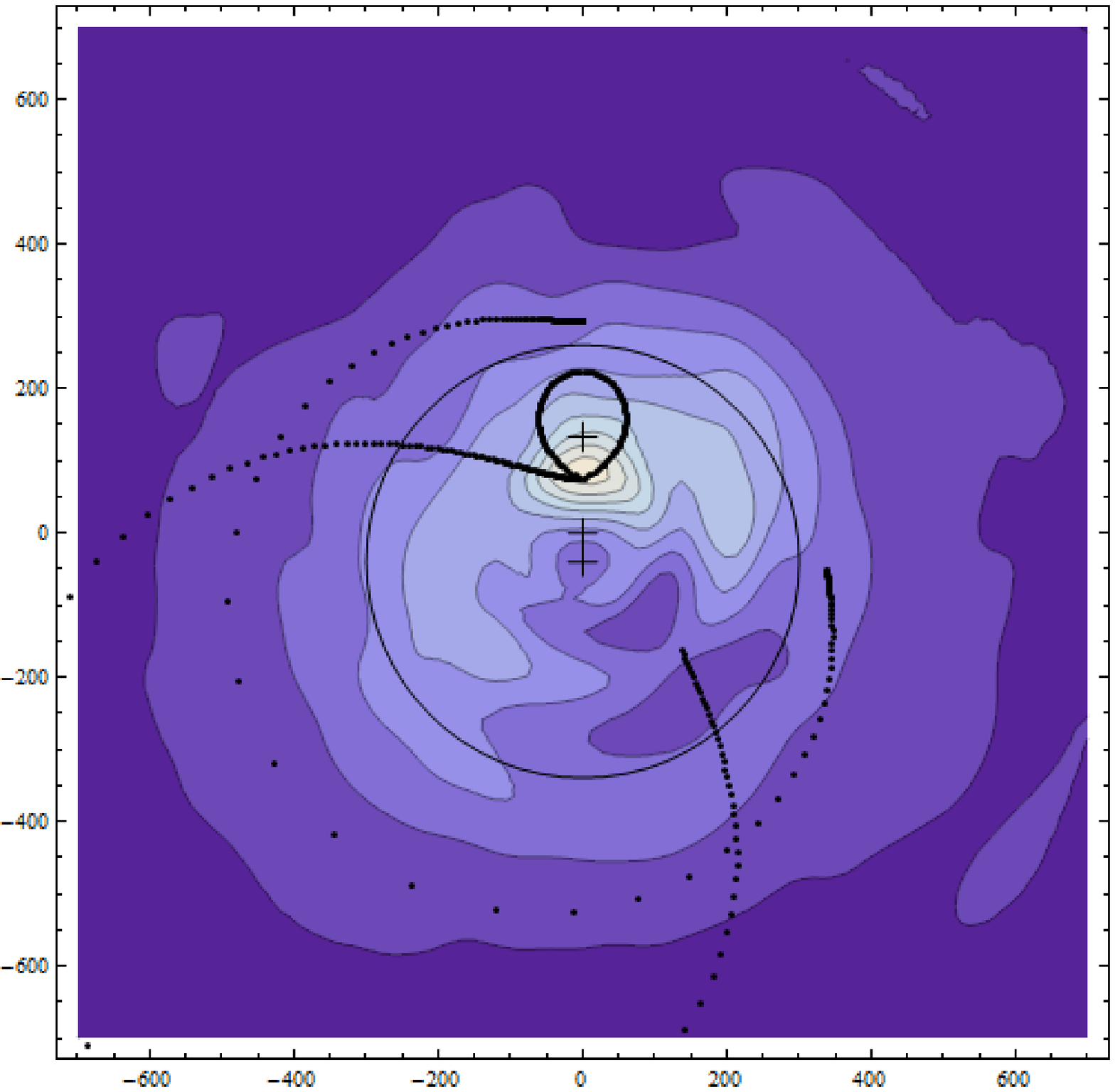}

\figcaption{Fourier-filtered doppler tomograms of the O\,VI line at 
1032\AA, corrected for 
interstellar absorption as described in the 
text. Top left: the trailed spectrogram. Top right: the trailed 
spectrogram, reconstructed from the tomogram. Bottom: 
the Doppler tomogram. As in Figure~\ref{opticaltomo}, the
locations in velocity space of the Roche lobe, the centers of mass of
the two stars, the gas stream and the
disk velocities along the stream, and the edge of the disk are indicated.\label{fig:tomo}}

\end{figure}

\begin{figure}
\plotone{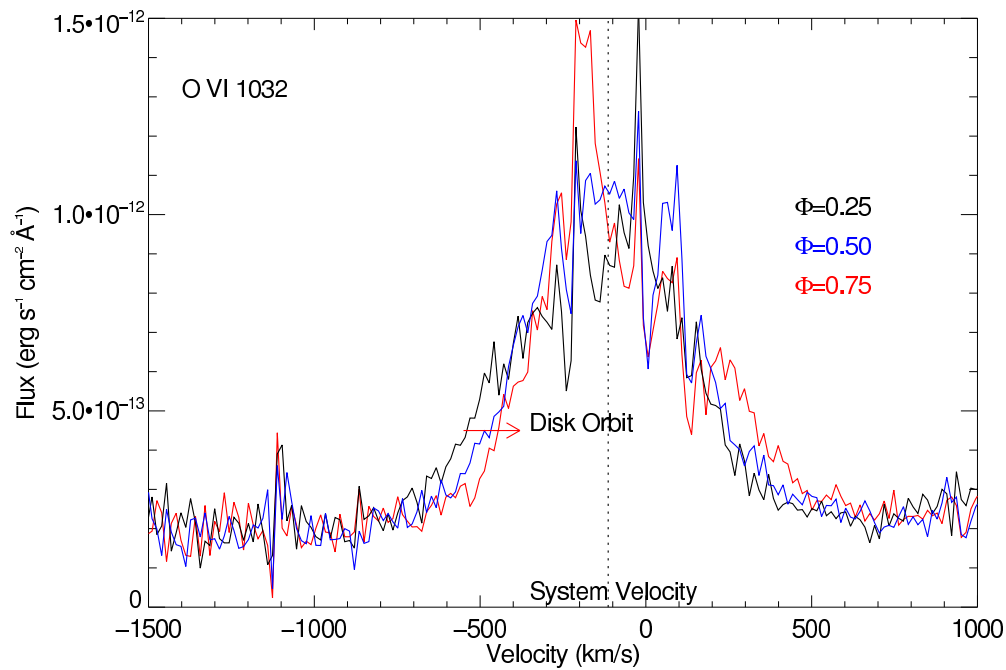}
\figcaption{A comparison of the O\,VI 1032\AA\ line at $\phi=0.25, 0.50,$ 
and 0.75. Each spectrum is the sum of 3 spectra at similar 
phases.\label{fig:lineav}}
\end{figure}

\begin{figure}
\plotone{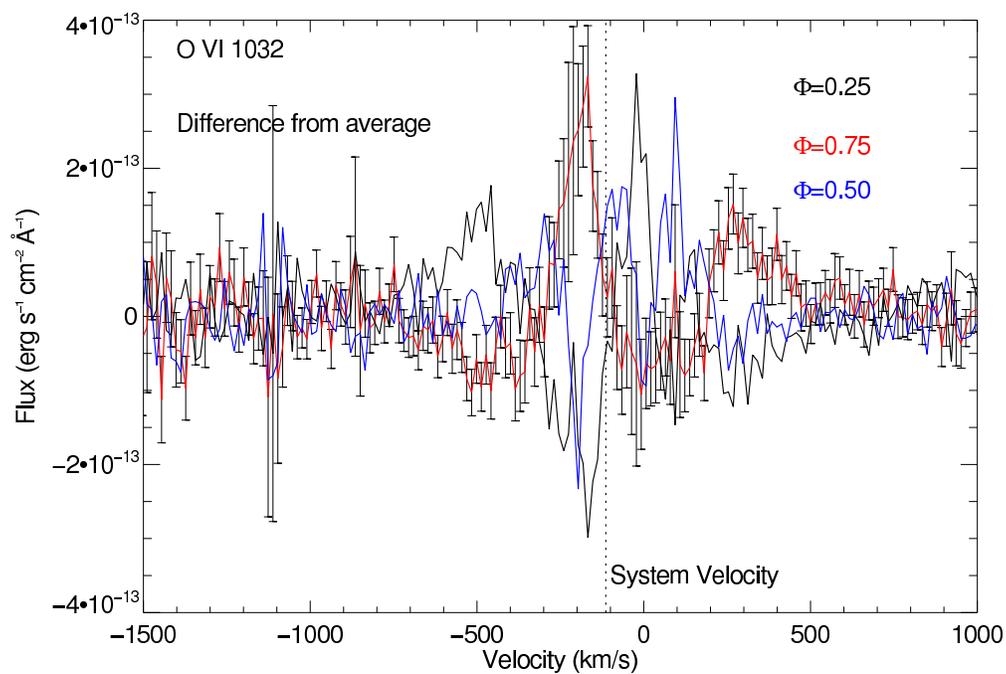}
\figcaption{A comparison of the differences of the O\,VI line from the 
orbital mean at $\phi=0.25$, 0.50, and 0.75. The error bars for one phase 
group are indicated.\label{fig:linedif}}

\end{figure}

\begin{figure}
\epsscale{0.8}
\plotone{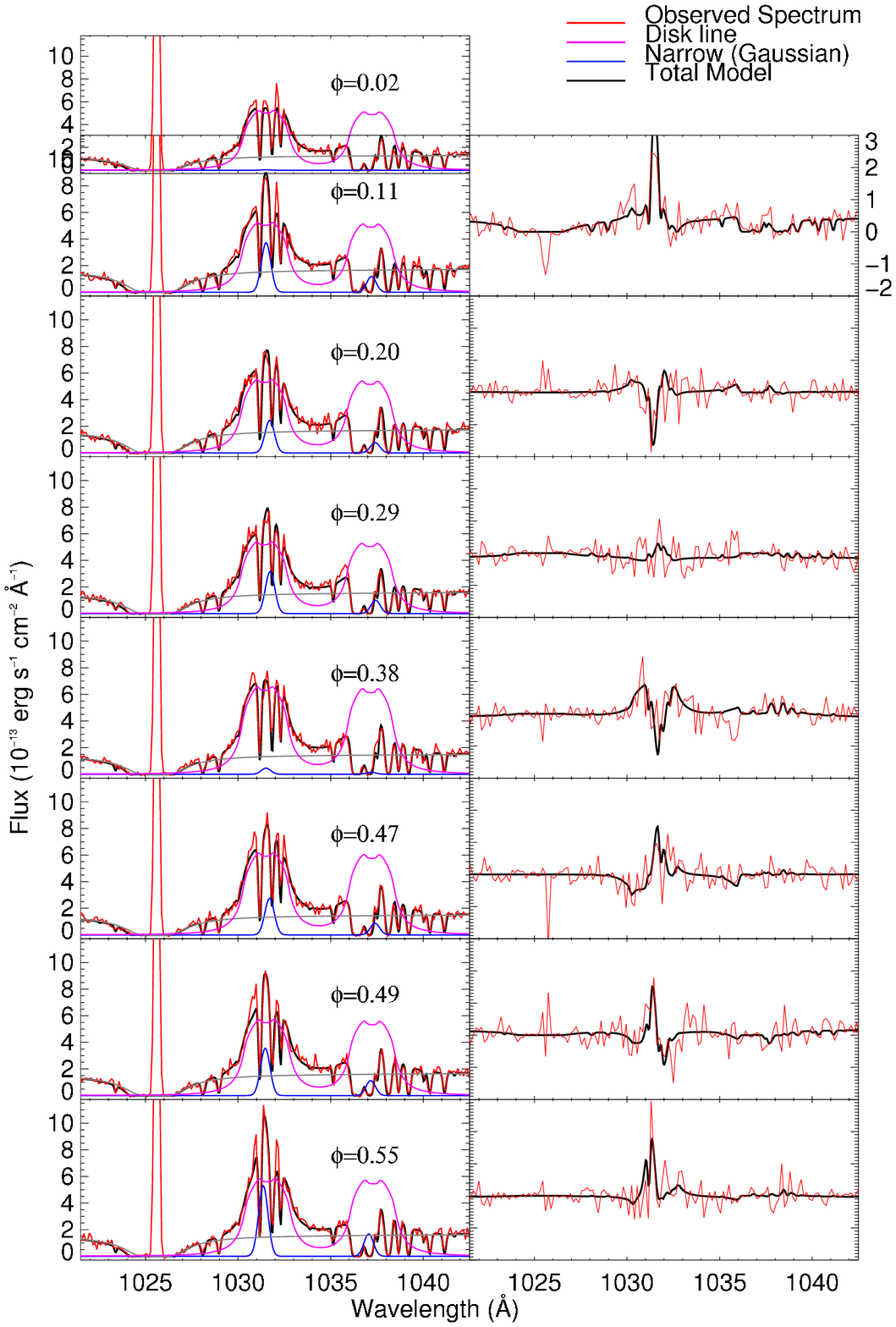}
\end{figure}

\newpage
\begin{figure}
\epsscale{0.7}
\plotone{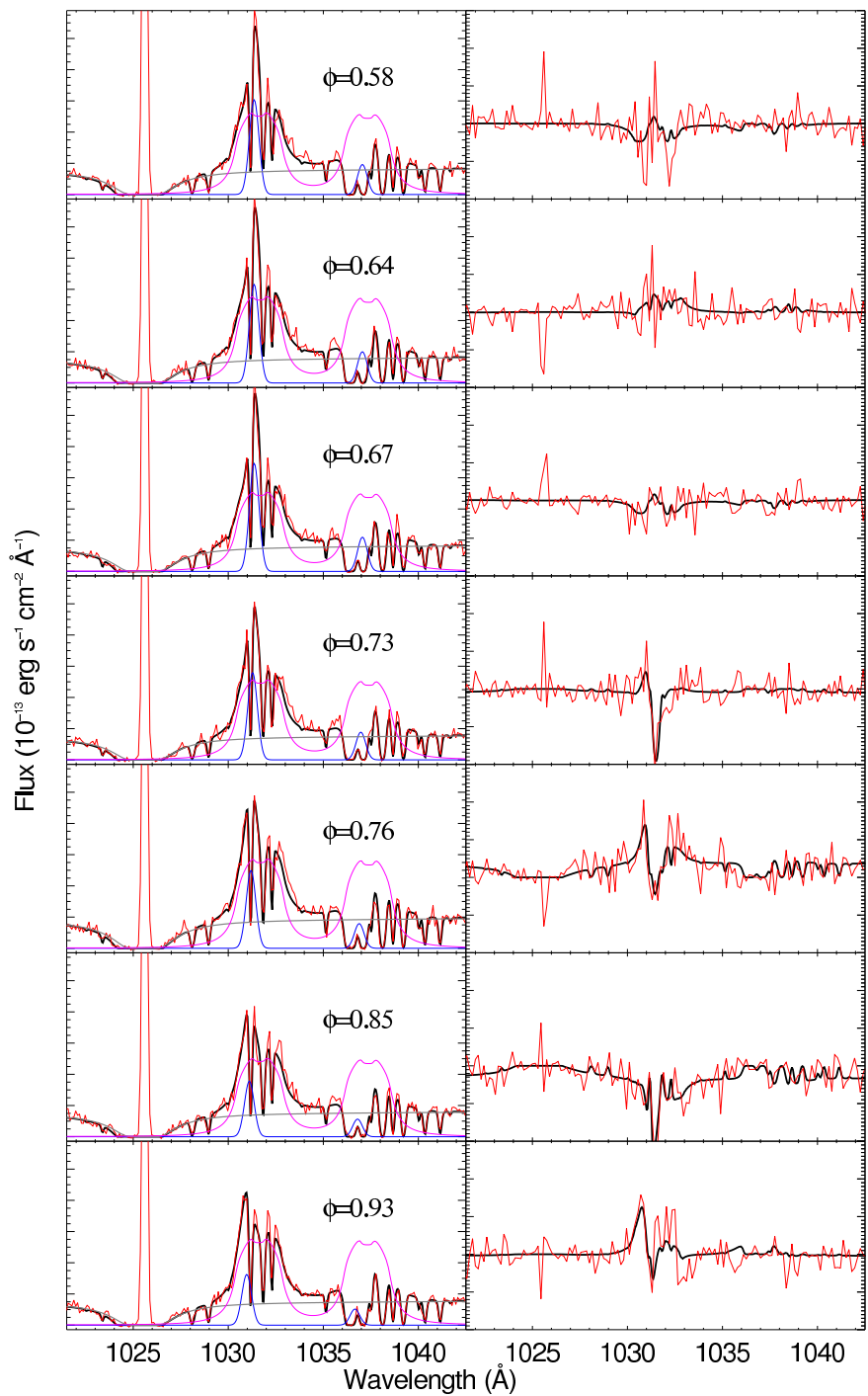}
\figcaption{A model of the O VI lines in Sco X-1 as the sum of a broad 
disk component and a narrow gaussian, with 
interstellar absorption applied. Left: the observed spectra with 
the model and model components. Right: the differences between successive 
observations and models, shown to highlight variability.
\label{fig:plotlines}} \end{figure}

\newpage
\begin{figure}
\plotone{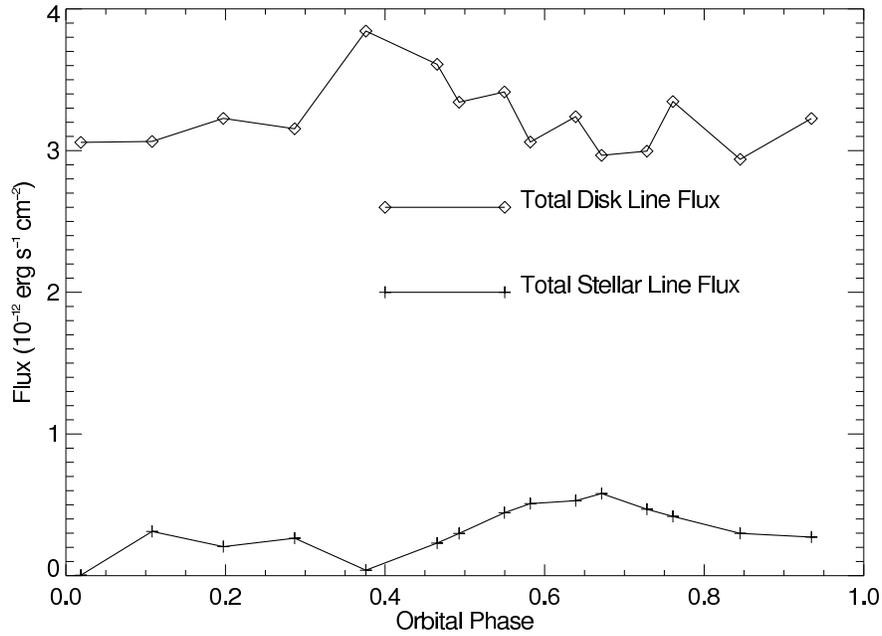}
\figcaption{The orbital variability in the flux in the disk and narrow line components to the O\,VI 
1032\AA\ emission line.}
\end{figure}

\newpage
\begin{figure}
\plotone{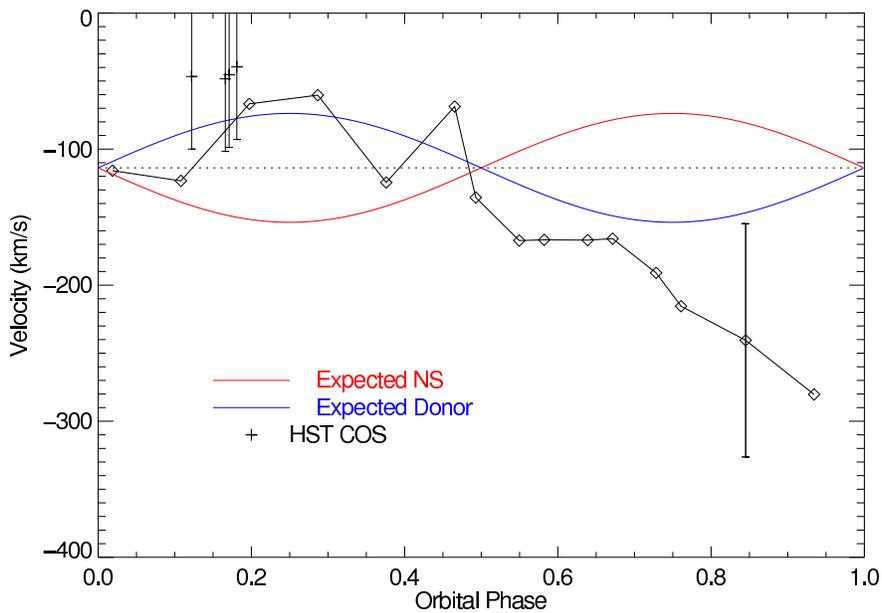}
\figcaption{The narrow line velocity versus orbital phase, as determined from fits to the O\,VI 
1032\AA\ emission line, along with HST COS fits to the O\,V 1371\AA\ 
line. The error bars show the Gaussian widths of the lines.}
\end{figure} 

\begin{figure}
\plotone{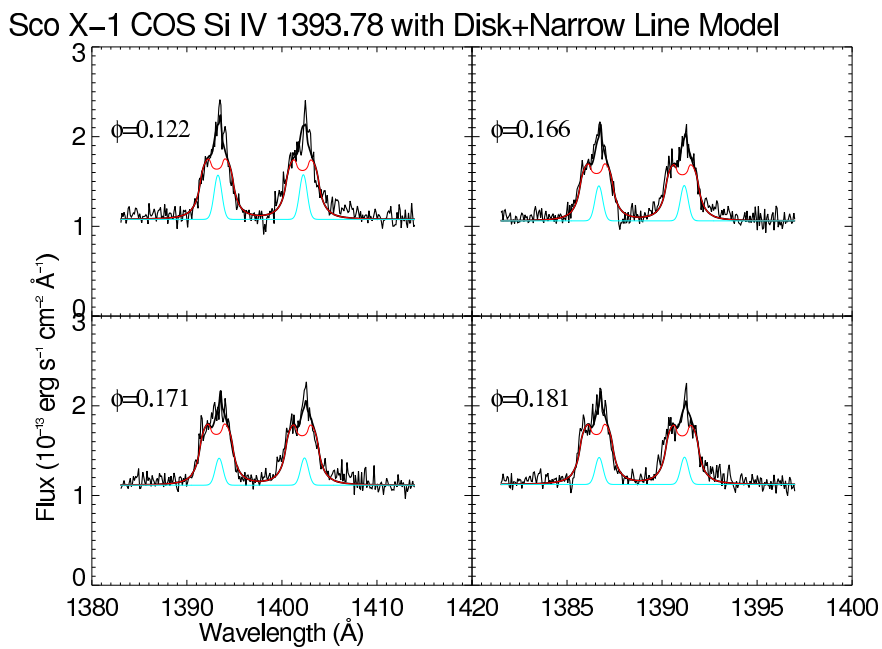}
\figcaption{The model of a disk line broadened by turbulence and Keplerian 
rotation, together with a narrow Gaussian line, applied to HST COS spectra 
of the Si\,IV doublet at 1393\AA\ and 1403\AA. \label{fig:silinefit}}
\end{figure}

\begin{figure}
\plotone{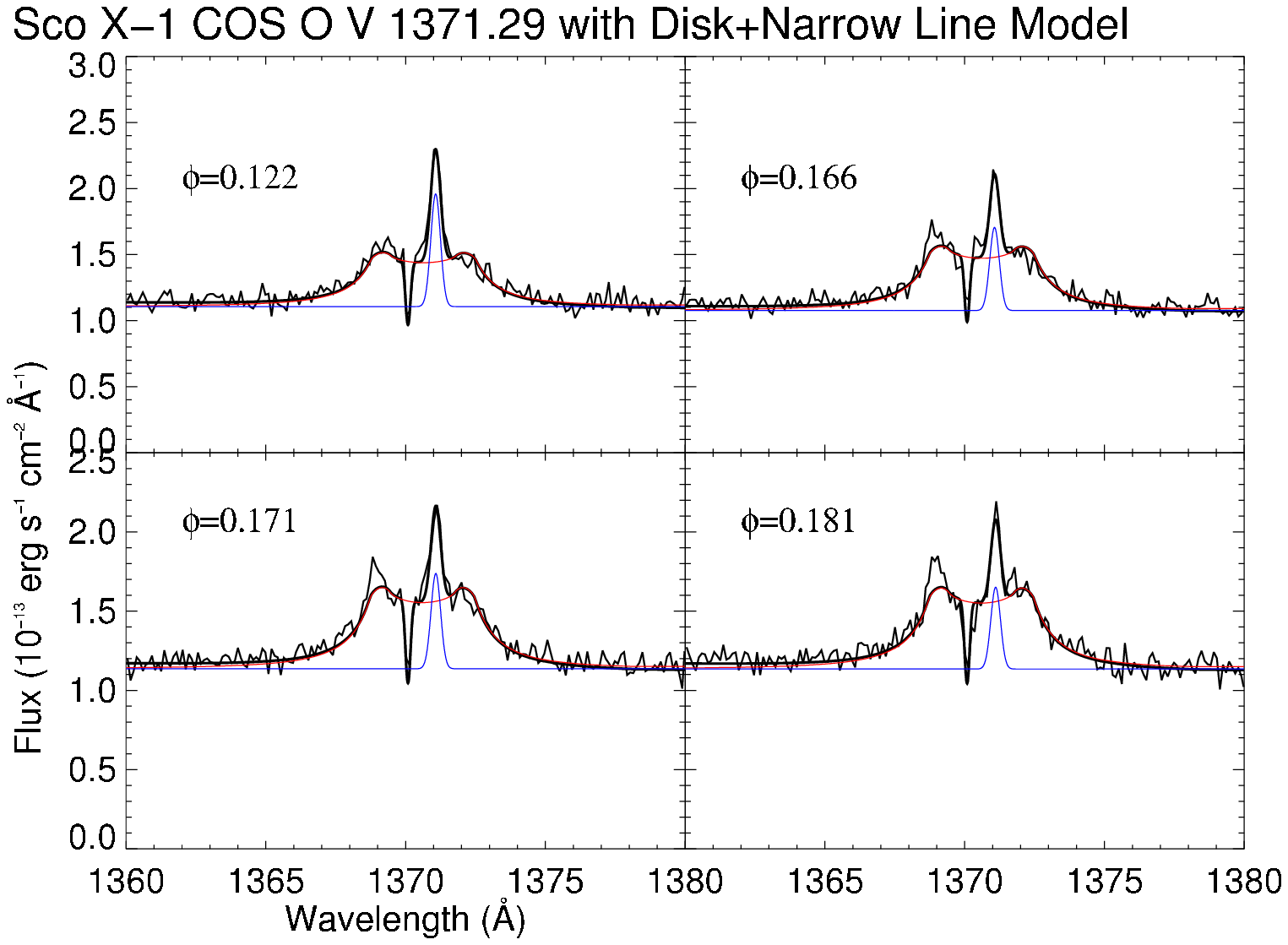}
\figcaption{The model of a disk line broadened by turbulence and Keplerian 
rotation, together with a narrow Gaussian line, applied to HST COS spectra 
of the O\,V 1371\AA\ line.
\label{fig:olinefit}}
\end{figure}


\clearpage

\begin{references}

Abgrall, H., Roueff, E., Launay, F., Roncin, J.Y., Subtil, J.L. 1993, 
A\&AS, 101, 273

Abgrall, H., Roueff, E., Launay, F., Roncin, J.Y., \&\ Subtil, J.L. 1993,
A\&AS, 101, 323

Boroson, B., Kallman, T., Vrtilek, S.D., Raymond, J., Still, M.,
Bautista, M., \&\ Quaintrell, H. 2000, ApJ, 529, 414

Boroson, B., Kallman, T., \&\ Vrtilek, S.D. 2001, ApJ, 562, 925

Boroson, B., Vrtilek, S.D., Raymond, J., \&\ Still, M. 2007, ApJ, 667, 
1087

Bowen, I.S. 1934, PASP, 46, 146

Bowen, I.S. 1935, ApJ, 81, 1

Cardelli, J.A., Clayton, G.C., \&\ Mathis, J.S. 1989, ApJ, 345, 245

Bradshaw, C.F., Fomalont, E.B., \&\ Geldzahler, B.J. 1999, ApJ, 512, L121

Cowley, A.P. \&\ Crampton, D. 1975, ApJ, 201, L65

Diplas, A., \&\ Savage, B.D. 1994, ApJ, 427, 274

Feldman, P.D., Sahnow, D.J., Kruk, J.W., Murphy, E.M., \&\ Moos, H.W. 
2001, JGR, 106, 8119

Galloway, D., Premachandra, S., Steeghs, D., Marsh, T., Casares, J., \&\ 
Cornelisse, R. 2014, ApJ, 781, 14

Gillmon, K., Shull, J.M., Tumlinson, J., \&\ Danforth, C. 2006, ApJ, 636, 
891

Gordon, K.D., Cartledge, S., \&\ Clayton, G.C. 2009, ApJ, 705, 1320

Gottlieb, E.W., Wright, E.L., \&\ Liller, W. 1975, ApJ, 195, L33

Horne, K. 1995, A\&A, 297, 273

Kallman, T., Boroson, B., \&\ Vrtilek, S.D. 1998, ApJ, 502, 441

Kallman, T.R., \&\ McCray, R. 1982, ApJS, 50, 263

Kallman, T.R., Raymond, J.C., \&\ Vrtilek, S.D. 1991, ApJ, 370, 717

Kupka, F., Piskunov, N., Ryabchikova, T.A., Stempels, H.C, \& Weiss, W.W.
1999, A\&AS, 138, 119

McCandliss, S.R. 2003, PASP, 115, 651

McCray, R., Wright, C., \&\ Hatchett, S. 1977, ApJ, 211, 29

Marsh, T.R., \&\ Horne, K. 1988, MNRAS, 235, 269

Marsh, T.R. 2005, A\&SS, 296, 403

Moos, H.W., Cash, W.C., Cowie, L.L., Davidsen, A.F., Dupree, A.K.,
Feldman, P.D., Friedman, S.D., Green, J.C., et al. 2000, ApJ, 
538, L1

Mu\~noz-Darias, T., Mart\`inez-Pais, I.G., Casares, J., Dhillon, V.S., Marsh, T.R., Cornelisse, R., 
Steeghs, D., \&\ Charles, P.A. 2007, MNRAS, 379, 1637

Nelder, J.A., \&\ Mead, R. 1965, CompJ, 7, 308

Osterman, et al. 2011. Ap\&SS, 335, 257

Raymond, J.C. 1993, ApJ, 412, 267

Robert, C., Pellerin, A., Aloisi, A., Leitherer, C., Hoopes, C., \&\
Heckman, T.M. 2003, ApJS, 144, 21

Sahnow, D.J., Moos, H.W., Ake, T.B., Andersen, J., Anderson, B.-G.,
Andre, M., Artis, D., Berman, A.F., Blair, W.P., Brownsberger, K.R., et 
al. 2000, ApJ, 538, L7

Sandage, A.R., et al. 1966, ApJ, 146, 316

Schachter, J., Filippenko, A.V., \&\ Kahn, S.M.  1989, ApJ,
340 1049

Sokolov, N.A. 2011, astro-ph/10107.1894

Steeghs, D., \&\ Casares, J. 2002, ApJ, 568, 273

Vrtilek, S.D., Raymond, J.C., Garcia, M.R., Verbunt, F., Hasinger,
G., \&\ Kurster, M. 1990, A\&A, 235, 162

Vrtilek, S.D., Penninx, W., Raymond, J.C., Verbunt, F., Hertz, P., Wood, 
K., Lewin, W.H.G., \&\ Mitsuda, K. 1991, ApJ, 376, 278

Willis, A.J. et al. 1980, ApJ, 237, 596

\end{references}
\end{document}